\documentclass[%
aapm,
mph,%
amsmath,amssymb,
reprint,%
]{revtex4-2}
\usepackage{graphicx}
\usepackage{dcolumn}
\usepackage{bm}
\usepackage{amsmath}
\usepackage{mathtools}
\usepackage{braket}
\usepackage{physics}
\usepackage{afterpage}
\usepackage{nicefrac}
\usepackage{xcolor}
\usepackage{amsfonts}
\usepackage{amssymb}
\usepackage{comment}

\usepackage[mathlines]{lineno}
\modulolinenumbers[5]
\linenumbers\relax 

\nolinenumbers

\newcommand{\JC}[1]{\textcolor{red}{#1}}

\begin{document}
	\date{\today}
	
	\title{Chirped Fractional Stimulated Raman Adiabatic Passage}
	
	\def\ARL{DEVCOM Army Research Laboratory, Adelphi, MD 20783}
	\def\UCBer{Department of Physics, University of California, Berkeley, CA 94720, USA}
	\def\Mainz{Johannes Gutenberg-Universit\"at Mainz, 55128 Mainz, Germany}
	\def\HIM{Helmholtz-Institut Mainz, GSI Helmholtzzentrum f{\"u}r Schwerionenforschung, 55128 Mainz, Germany}
	\def\Stevens{Department of Physics, Stevens Institute of Technology, Hoboken, NJ 07030}
	\def\Darmstadt{Technische Universit\"at Darmstadt, 
		Institut f\"ur Angewandte Physik, 
		D-64289 Darmstadt, Germany}
	

	\author{~Jabir~Chathanathil}
	\affiliation{\Stevens}
	\affiliation{\ARL}

\author{Aneesh Ramaswamy}
\affiliation{\Stevens}

\author{Vladimir~S.~Malinovsky}
\affiliation{\ARL} 

\author{Dmitry Budker}
\affiliation{\Mainz, \HIM}
\affiliation{\UCBer}

\author{Svetlana~A.~Malinovskaya}
\affiliation{\Stevens, \HIM}
\affiliation{\Darmstadt}

\begin{abstract}

Stimulated Raman Adiabatic Passage (STIRAP) is a widely used method for adiabatic population transfer in a multilevel system. In this work, we study STIRAP under novel conditions and focus on the fractional, F-STIRAP, which is known to create a superposition state with the maximum coherence. In both configurations, STIRAP and F-STIRAP, we implement pulse chirping aiming at a higher contrast, a broader range of parameters for adiabaticity, and enhanced spectral selectivity. Such goals target improvement of quantum imaging, sensing and metrology, and broaden the range of applications of quantum control techniques and protocols. In conventional STIRAP and F-STIRAP, two-photon resonance is required conceptually to satisfy the adiabaticity condition for dynamics within the dark state. Here, we account for a non-zero two-photon detuning and present control schemes to achieve the adiabatic conditions in STIRAP and F-STIRAP through a  skillful compensation of the two-photon detuning by pulse chirping. We show that the chirped configuration – C-STIRAP – permits adiabatic passage to a predetermined state among two nearly degenerate final states, when conventional STIRAP fails to resolve them.  We demonstrate such a selectivity within a broad range of parameters of the two-photon detuning and the chirp rate. In the C-F-STIRAP, chirping of the pump and the Stokes pulses with different time delays permits a complete compensation of the two-photon detuning and results in a selective maximum coherence of the initial and the target state with higher spectral resolution than in the conventional F-STIRAP.

\end{abstract}

\maketitle

\section{Introduction}

Since its discovery in 1990, Stimulated Raman Adiabatic Passage (STIRAP) has developed into a prominent method of quantum coherent control \cite{STIRAP_Original_1990}.  
Owing to its robustness, STIRAP has been used in a vast variety of research fields as detailed in the ``roadmap" paper published in 2019 \cite{Roadmap}. The applications of STIRAP continue to extend, advancing state control in solid state materials, e.g., NV centers \cite{Florian}, and SiV centers in diamond \cite{Becher}, creating ultracold molecules using mixed intermediate states \cite{Bl},  performing geometric gates in superconducting qubits by implementing shortcut-to-adiabaticity \cite{PRApplied2023}, mastering nuclear coherent population transfer to the $^{229m}$Th isomer using x-ray pulses \cite{Adriana},  
efficiently swapping population using an arbitrary initial state \cite{SWAP}, 
designing a digitized version of STIRAP \cite{Ma22}, 
and imaging stars via quantum communication techniques \cite{StarPRL2022}.  

An extension of STIRAP - the fractional STIRAP (F-STIRAP) - may prove useful for imaging, sensing and detection by virtue of the generation of an enhanced signal as well as signal sustainability upon propagation through a medium.
F-STIRAP generates  a coherent superposition of the initial and the final states by manipulating the duration of the Stokes pulse, which vanishes simultaneously with the pump pulse \cite{F-STIRAP_1999}. F-STIRAP was applied in experiments in Rb atomic vapor  to maximize atomic coherence, which led to the enhancement of coherent Raman scattering \cite{Scully_F-STIRAP}. The practicality of this method is based on a relative flexibility of the key control parameters relevant for both  STIRAP and F-STIRAP such as the fields strength, the ratio of the pump to the Stokes Rabi frequency, the Stokes-pump pulse delay and  
the pulse duration. 
Besides, chirping of ultrafast pulses brings  spectroscopic advantages as shown in a number of papers \cite{Ma_2007,Budker_2016,Pandya_2020,Ch_2021}. In  \cite{MagnesPRA94}, chirped STIRAP permitted the selective excitation of two nearly degenerate states by changing the sign of the chirp. However, F-STIRAP has never been investigated to improve the spectral resolution of imaging and detection techniques. This motivated us to explore thoroughly the effects of chirping pulses in STIRAP as well as F-STIRAP processes. We show that chirping both pulses with equal rates in C-STIRAP is useful to satisfy the conditions for adiabaticity even in the presence of the two-photon detuning, which is known to be responsible for non-adiabatic coupling in the conventional STIRAP. 
As s result, in a nearly degenerate four-level system, the population can be driven to a desired level by controlling the sign of the chirp of the Stokes and the pump pulses. Chirped Adiabatic Passage (CHIRAP) has been used for selective population transfer to one of the fine structure states in Na vapor \cite{Melinger_1992, Melinger_1994}. It differs from C-STIRAP in that there is no temporal delay between the pulses in CHIRAP. In C-F-STIRAP, chirping of the pulses results in an improved sensitivity of the selective creation of maximum coherence between the initial and final levels.

\begin{figure}
\includegraphics[scale=0.48]{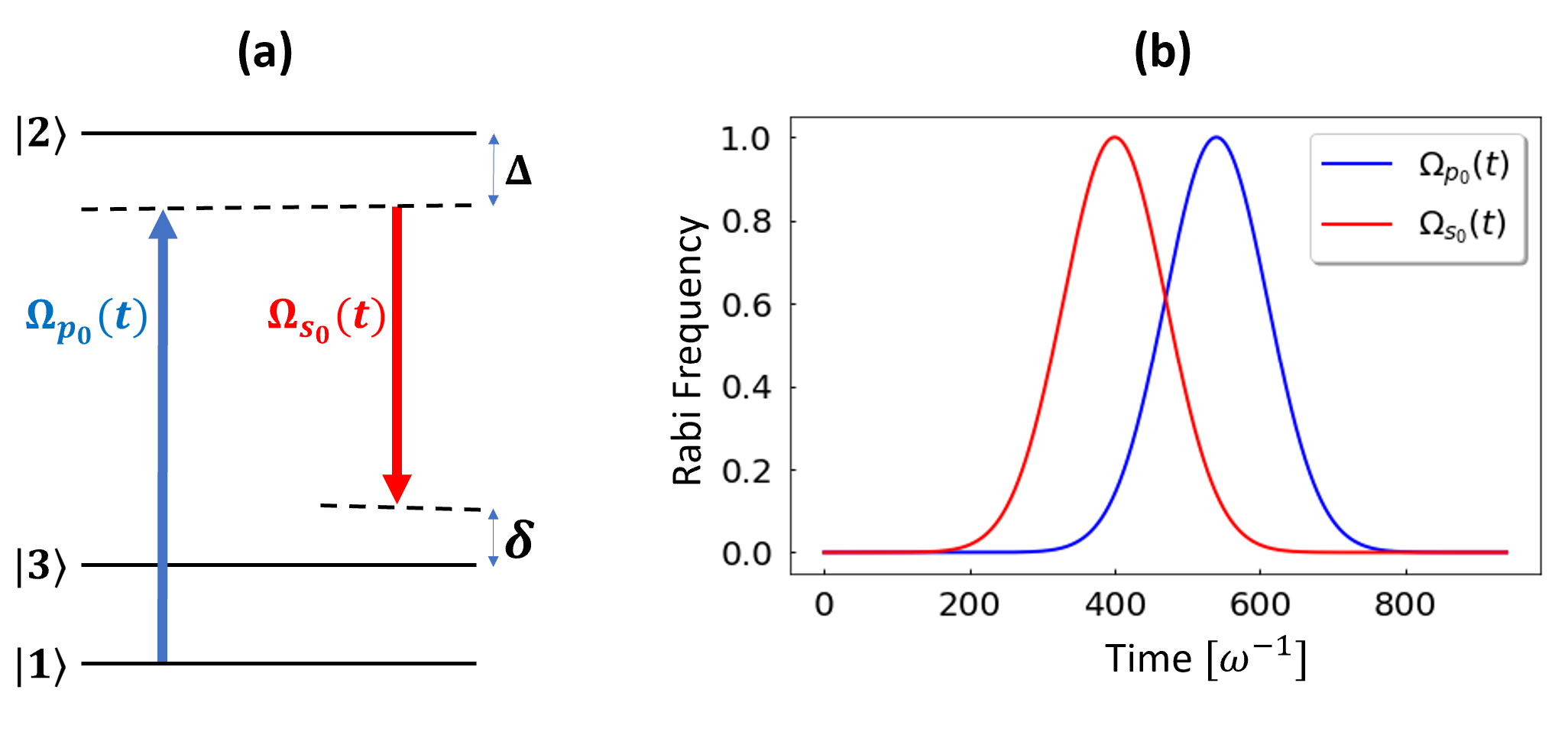}
\caption{The coupling scheme of STIRAP in the three-level $\lambda$ system.  The Rabi frequencies of the pump and the Stokes pulses are shown in (b). The one-photon and the two-photon detunings are defined as   $\Delta=\omega_2-\omega_1-\omega_p$ and $\delta=\omega_p-\omega_s-(\omega_3-\omega_1)$ respectively.}\label{3-level}
\end{figure}

The paper is organized as follows. First, we discuss the configuration of C-STIRAP in a three-level $\lambda$ system and then explain the selective excitation of nearly degenerate final states through dressed state analysis; 
next, we address the C-F-STIRAP and show how the mechanism of selective excitation has modified for the case of the final coherent superposition state. 

\section{CHIRPED STIRAP (C-STIRAP)}

\subsection{C-STIRAP in a three-level $\lambda$ system}
In conventional STIRAP, the pump and the Stokes pulse interact with a three-level system making a complete and adiabatic population transfer from the initial to the final state. The schematic of such a process in a $\lambda$ system is shown in Fig. \ref{3-level}(a), where $\Omega_{p_0}(t)$ and $\Omega_{s0}(t)$ are the Rabi frequency of the pump and the Stokes pulses respectively. The one-photon and two-photon detunings are  $\Delta=\omega_2-\omega_1-\omega_p$, and $\delta=\omega_p-\omega_s-(\omega_3-\omega_1)$. The counter-intuitive sequence and a significant overlap of the two pulses, as shown in Fig. \ref{3-level}(b), are required for keeping the STIRAP dynamics adiabatic. Another requirement for the STIRAP adiabaticity is the presence of the two-photon resonance  \cite{Vitanov_beyond_resonance_2010}.
Here, we demonstrate that it is possible to bypass this requirement by chirping the input pulses.

The Hamiltonian of the three-level $\lambda$ system interacting with the linearly chirped pump and Stokes pulses has the form
\begin{eqnarray}\label{Hamil-Schrodinger}
\mathbf{H}(t) =
\sum_{i=1}^{3}\hbar \omega_i \ket{i} \bra{i} - \left(\mu_{21}E_p(t)\ket{1}\bra{2} + h.c.\right) \nonumber \\ -\left(\mu_{23}E_s(t)\ket{3}\bra{2} + h.c.\right),
\end{eqnarray}
where the pump and the Stokes fields $E_{p,s}(t)$ having Gaussian envelopes and carrier frequencies $\omega_{p,s}$, chirp rates $\alpha_{p,s}$, and central times $t_{p,s}$ are given by:
\begin{gather}\label{C-STIRAP_pulses}
\scalebox{0.9}{$E_{p,s}(t)=E'_{p_0,s_0}  e^{\frac{-(t-t_{p,s})^2}{\tau_{p,s}^2}} \cos[\omega_{p,s}(t-t_{p,s})+\tfrac{\alpha_{p,s}}{2}(t-t_{p,s})^2].$}
\end{gather}
The dynamics of the system is governed by the Schr\"odinger equation
\begin{equation}\label{Schrodinger}    i\hbar\frac{\partial}{\partial t}\ket{\psi(t)} =\mathbf{H}(t)\ket{\psi(t)}\,, \ \ \ \ket{\psi(t)} = \sum_{n=1}^{3}a_n(t)\ket{n}. \end{equation}

Using the transformations
\begin{equation}\label{STIRAP-tranformations}
\begin{aligned}
a_1 (t) &= \tilde{a_1}(t) e^{i\omega_p(t-t_p) + \frac{i}{2}\alpha_p(t-t_p)^2} \\
a_2 (t) &= \tilde{a_2} (t) \\
a_3 (t) &=\tilde{a_3}(t) e^{i\omega_s(t-t_s) + \frac{i}{2}\alpha_s(t-t_s)^2}
\end{aligned}
\end{equation}
and applying the rotating wave approximation leads to the field-interaction Hamiltonian 
\begin{eqnarray}\label{Ham3level-C-STIRAP}
\mathbf{H}(t) =
\frac{\hbar}{2} \left( \begin{array}{cccc} 0   & \Omega_{p_0} (t)    &   0\\
\Omega_{p_0} (t) &  2\Delta(t) &  \Omega_{s_0}(t) \\
0  &  \Omega_{s_0}(t)&  2\delta(t)\\
\end{array} \right), 
\end{eqnarray}
where the Rabi frequencies $\Omega_{p_0} (t)=-\mu_{21} E_{p_0}(t)/\hbar$ and $\Omega_{s_0} (t)= -\mu_{32} E_{s_0}(t)/\hbar$ are real quantities with the Gaussian amplitudes $E_{p_0, s_0}(t)=E_{p_0,s_0}  \exp{-(t-t_{p,s})^2 / \tau_{p,s}^2}$ and the chirp rates are relabeled as $\alpha_p=\alpha$ and $\alpha_s=\beta$. We define the time-dependent one-photon and two-photon detunings as $\Delta(t)=\Delta-\alpha(t-t_p)$ and $\delta(t)=-\delta+\beta(t-t_s)-\alpha(t-t_p)$ respectively. In  this section, the value of peak Rabi frequencies and time durations are taken to be $\Omega_{p_0, s_0}$ = 1.0[$\omega$] and $\tau_{p,s}=100[\omega^{-1}]$.

\begin{figure}  \includegraphics[scale=0.48]{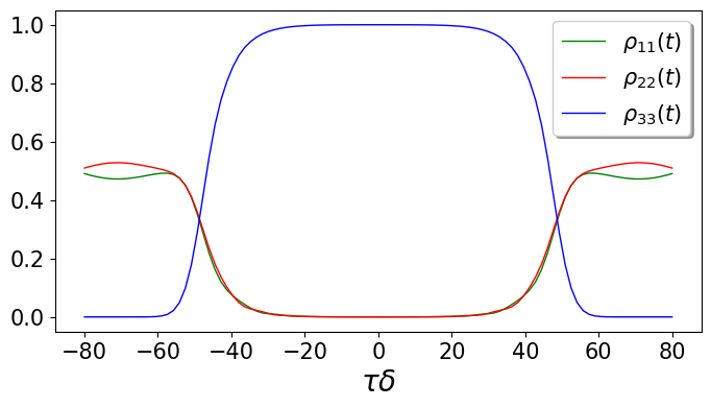}
\caption{The state populations in STIRAP as a function of $\tau \delta$ when the pulses are not chirped. The population is fully transferred to the final state in the vicinity of the two-photon resonance $\delta=0$. Here, $\tau=100[\omega^{-1}]$.}\label{STIRAP_detuning_vs_populations}
\end{figure}

The state populations at $t\rightarrow \infty$ as a function of $\tau \delta$ are given in Fig. \ref{STIRAP_detuning_vs_populations}. A full population transfer to the final state occurs in the vicinity of two-photon resonance. For non-zero values of $\delta$, the evolution of populations is not adiabatic as demonstrated in the next subsection. However, when the pump and Stokes pulses are chirped with carefully chosen chirp rates, the population can be adiabatically transferred even in the presence of the two-photon detuning. A closer look at the dressed state picture in C-STIRAP reveals the conditions for adiabaticity when $\delta \neq 0$.

\subsection{Dressed state analysis of the C-STIRAP:  A three-level $\lambda$ system}
Consider 
a unitary rotation matrix $\textbf{T}(t)$,
\begin{equation}\label{Tmatrix}
\textbf{T}(t) = 
\begin{pmatrix}
\sin\theta(t) \sin\phi(t)	&	\cos\theta(t)	&	\sin\theta(t) \cos\phi(t)\\
\cos\phi(t)	&	0 	&	-\sin\phi(t)\\
\cos\theta(t) \sin\phi(t)	&	-\sin\theta(t)	&	\cos\theta(t) \cos\phi(t)\\
\end{pmatrix}
\end{equation}
with mixing angles defined by:
\begin{eqnarray}
\tan\theta(t) &=& \frac{\Omega_{p_0}(t)}{\Omega_{s0}(t)}, \\ \nonumber	\tan2\phi(t) &=& \frac{\sqrt{|\Omega_{p_0}(t)|^2+|\Omega_{s0}(t)|^2}}{\Delta(t)}\,.
\end{eqnarray}
Rotation of the amplitudes 
$\tilde{\textbf{a}}(t)$ using this matrix $\mathbf{T(t)}$ matrix, gives the dressed state  basis $\bf{c_d}(t)=\textbf{T}^\dagger(t) \tilde{\textbf{a}}(t)$; in this basis the dressed state Hamiltonian, $\textbf{H}_d(t)$, reads
\begin{widetext}
\begin{eqnarray}\label{H_dressed}
\mathbf{H}_d(t)
&=& \mathbf{T^ \dagger}(t)\mathbf{H}(t)\mathbf{T}(t)-i \hbar \textbf{T}^\dagger(t)\dot{\textbf{T}}(t) \nonumber \\
&=&\hbar
\begin{pmatrix}
\lambda_+(t)	&	0	&	0\\
0	&	\lambda_0(t) 	&	0\\
0	&	0	&	\lambda_-(t) 
\end{pmatrix}
+ \hbar \delta(t) \cos^2\theta(t)
\begin{pmatrix}
\sin^2\phi(t)	&	-\tan\theta(t) \sin\phi(t)	&	(1/2)\sin 2\phi(t)\\
-\tan\theta(t) \sin\phi(t)	&	\tan^2\theta(t) 	&	-\tan\theta(t) \sin\phi(t)\\
(1/2)\sin 2\phi(t)	&	-\tan\theta(t) \sin\phi(t)	&	\cos^2\phi(t)
\nonumber
\end{pmatrix}\\
&-& i\hbar
\begin{pmatrix}
0	&	-\dot{\theta}(t) \sin \phi(t) 	&	-\dot{\phi}(t)\\
\dot{\theta}(t) \sin \phi(t)	&	0 	&	\dot{\theta}(t) \cos \phi(t) \\
\dot{\phi}(t)	&	-\dot{\theta}(t) \cos \phi(t)	&	0\\
\end{pmatrix}, 
\end{eqnarray}
\end{widetext}
where 
\begin{gather}\label{dressed_energies}
\scalebox{0.9}{$\begin{aligned}
\lambda_+(t) &= \frac{1}{2}\left(\Delta(t)+\sqrt{\left(\Delta(t)\right)^2 + |\Omega_{p_0}(t)|^2+|\Omega_{s_0}(t)|^2}\right) \\
\lambda_0(t) &= 0 \\ 
\lambda_-(t) &= \frac{1}{2}\left(\Delta(t)-\sqrt{\left(\Delta(t)\right)^2 + |\Omega_{p_0}(t)|^2+|\Omega_{s_0}(t)|^2}\right)
\end{aligned}$}
\end{gather}

For the process to be adiabatic, the dressed state Hamiltonian $\mathbf{H}_d(t)$ needs to be diagonal. The non-adiabatic contribution due to the second term in Eq. \eqref{H_dressed} is cancelled out by imposing the time-dependent two-photon detuning $\delta(t)$ to be zero. This can be done by choosing the chirp rates such that $-\delta+(\beta-\alpha)t+\alpha t_p-\beta t_s = 0 $. If $\alpha$ and $\beta$ are chosen to be equal, this condition becomes  $ \alpha(t_p-t_s)=\delta$ and a proper choice of $t_p, t_s$ and $\alpha$ eliminates this non-adiabatic term.

The third term in Eq. \eqref{H_dressed} constitutes the non-adiabatic contribution from $\textbf{T}^\dagger(t)\dot{\textbf{T}}(t)$, 
with
\begin{equation}
\dot{\theta}(t)= \frac{\Omega_{s_0}(t)\dot{\Omega}_{p_0}(t)-\Omega_{p_0}(t)\dot{\Omega}_{s_0}(t)}{\left(\Omega_{p_0}^2(t)+\Omega^2_{s_0}(t)\right)^{3/2}}\,.
\end{equation}
For adiabatic passage, the contribution from the third term must be negligible, requiring  $|\Dot{\theta}(t)|, |\dot{\phi}(t)|\ll|\lambda_{\pm}(t)|$. The conditions for mixing angles $\theta(t)$ and $\phi(t)$ are met in the presence of a significant overlap between the Stokes and the pump pulses with $\theta(t)$ and $\phi(t)$ varying very slowly.


When these adibaticity conditions are satisfied, 
the dressed state  
having the zero energy, $\lambda_0(t)=0$, is     \begin{equation}\label{dark_state}
\ket{\lambda_0(t)} = \cos\theta(t) \ket{\tilde{1}} - \sin\theta(t) \ket{\tilde{3}}.
\end{equation}
This state, known as `the dark state', smoothly evolves from initial bare state $\ket{\tilde{1}}$ to the final bare state $\ket{\tilde{3}}$ without having any component of the intermediate state $\ket{\tilde{2}}$. For the remainder of this section, the `tilde' on the wavefunction and eignenstates are dropped for convenience.

\begin{figure*}
\includegraphics[scale=0.55]{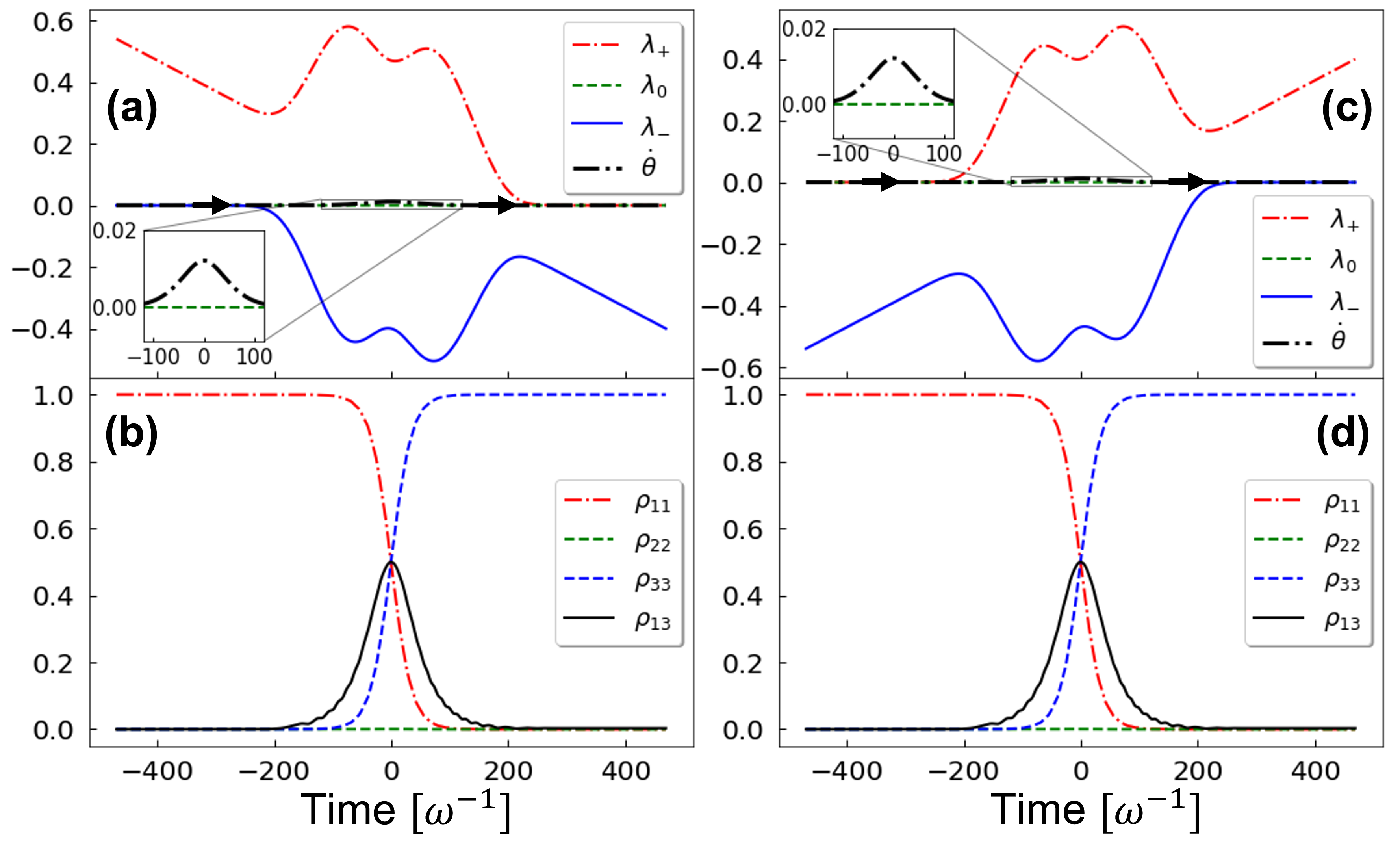}
\caption{The evolution of the dressed state energies and populations in the C-STIRAP when  $\delta > 0$ and $\alpha>0$, (a) and (b); and  $\delta < 0$ and $\alpha<0$, (c) and (d). The population is adiabatically transferred from the initial to the final state in both cases, owing to the choice of the chirp rates satisfying $\delta(t)=-\delta+\alpha(t_p-t_s)=0$, where $\alpha=\beta$. The non-adiabatic coupling parameter $V_{0\pm}$ is non-zero for a small duration when $|\lambda_+-\lambda_-|$ is the highest, as shown in the inset, implying the adiabatic nature of interaction. 
Here $\delta=0.14[\omega]$ and $\alpha= 1\times10^{-3} [\omega^2]$ in (a) and (b), and $\delta=-0.14[\omega]$ and $\alpha=-1\times10^{-3} [\omega^2]$ in (c) and (d). Other parameters are: $\Delta=0$, $t_s=-70[\omega^{-1}]$, $t_p=70[\omega^{-1}]$, $ \tau_{p,s}=100[\omega^{-1}]$ and peak Rabi frequencies are $\Omega_{p_0,s_0}=1.0[\omega]$.}\label{C-STIRAP_dressed}
\end{figure*}

Adiabatic population transfer in the presence of non-zero two-photon detuning is demostrated in Fig. \ref{C-STIRAP_dressed}. Here, the evolution of the dressed state energies $\lambda_{0,\pm}$, the non-adiabatic coupling parameter $V_{0\pm}$, the state populations and coherence 
are given for the chirp rates satisfying the condition $\delta(t)=0$. In (a) and (b), $\delta$ and $\alpha$ are positive, and in (c) and (d), these parameters  are negative. In both cases, the two-photon detuning $\delta$ is compensated by $\alpha(t_p-t_s)$ and the system dynamics is always aligned with the dark state having energy $\lambda_0 (t)=0$; the population is completely transferred adiabatically from the initial bare state $|1\rangle$ to the final bare state $|3\rangle$ without populating the intermediate state $|2\rangle$.

\begin{figure}
\includegraphics[scale=0.55]{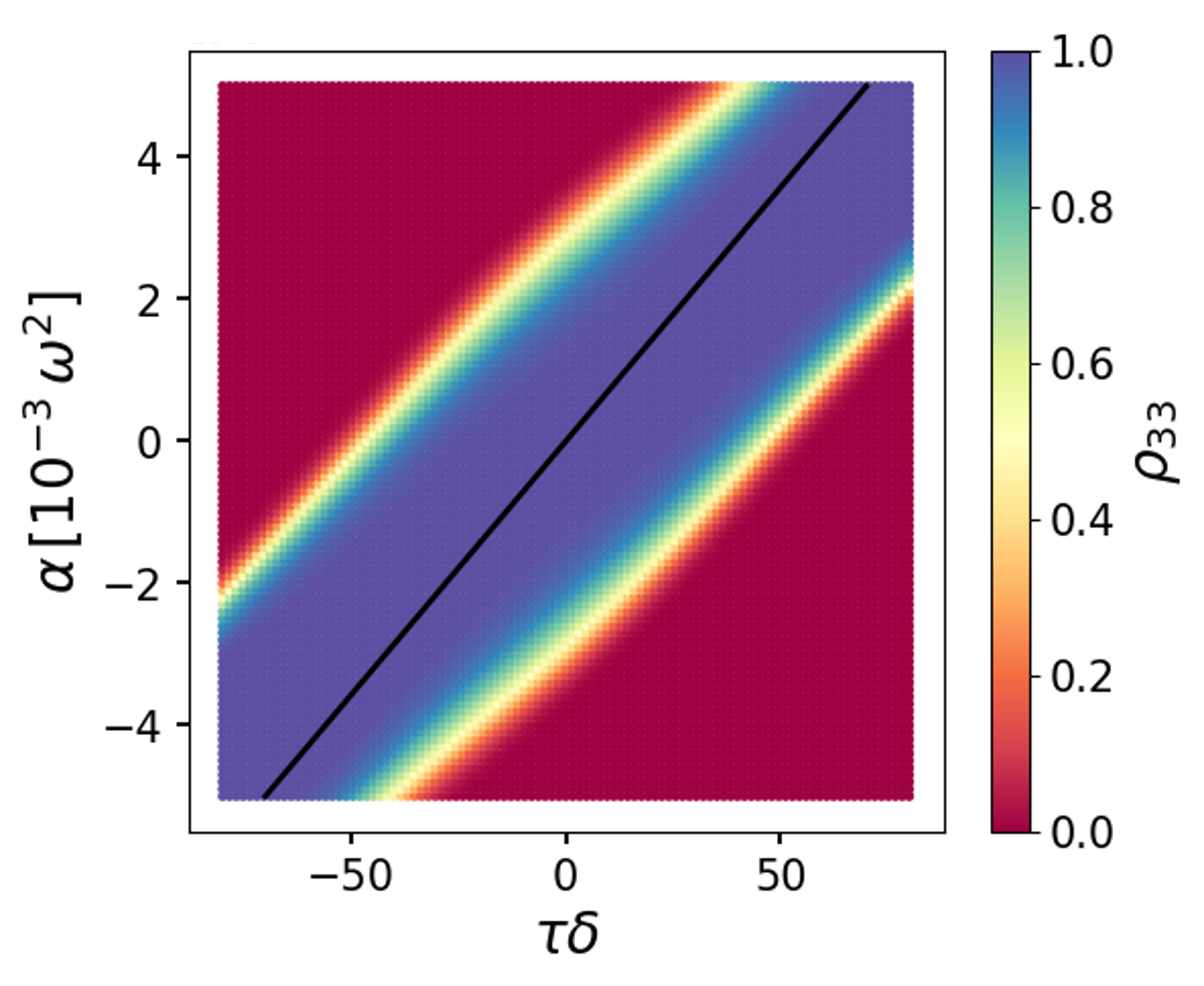}
\caption{Final state population in C-STIRAP as a function of two-photon detuning and chirp rate. The adiabatic transfer of population is achieved in the vicinity of the dark line satisfying $\alpha=\delta/(t_p-t_s)$, corresponding to $\delta(t)=0$.}\label{rho33_delta_vs_chirp_3-level}
\end{figure}

The robustness of this population is demonstrated in Fig. \ref{rho33_delta_vs_chirp_3-level} where final state population $\rho_{33}$ is plotted as a function of two-photon detuning $\delta$ and chirp rate $\alpha$. A broad area in the vicinity of the dark line satisfying the condition $\alpha=\delta/(t_p-t_s)$ indicates the robustness of this scheme.

\subsection{C-STIRAP in a four-level $\lambda$ system with two energetically close final states}

\begin{figure}	\includegraphics[scale=0.6]{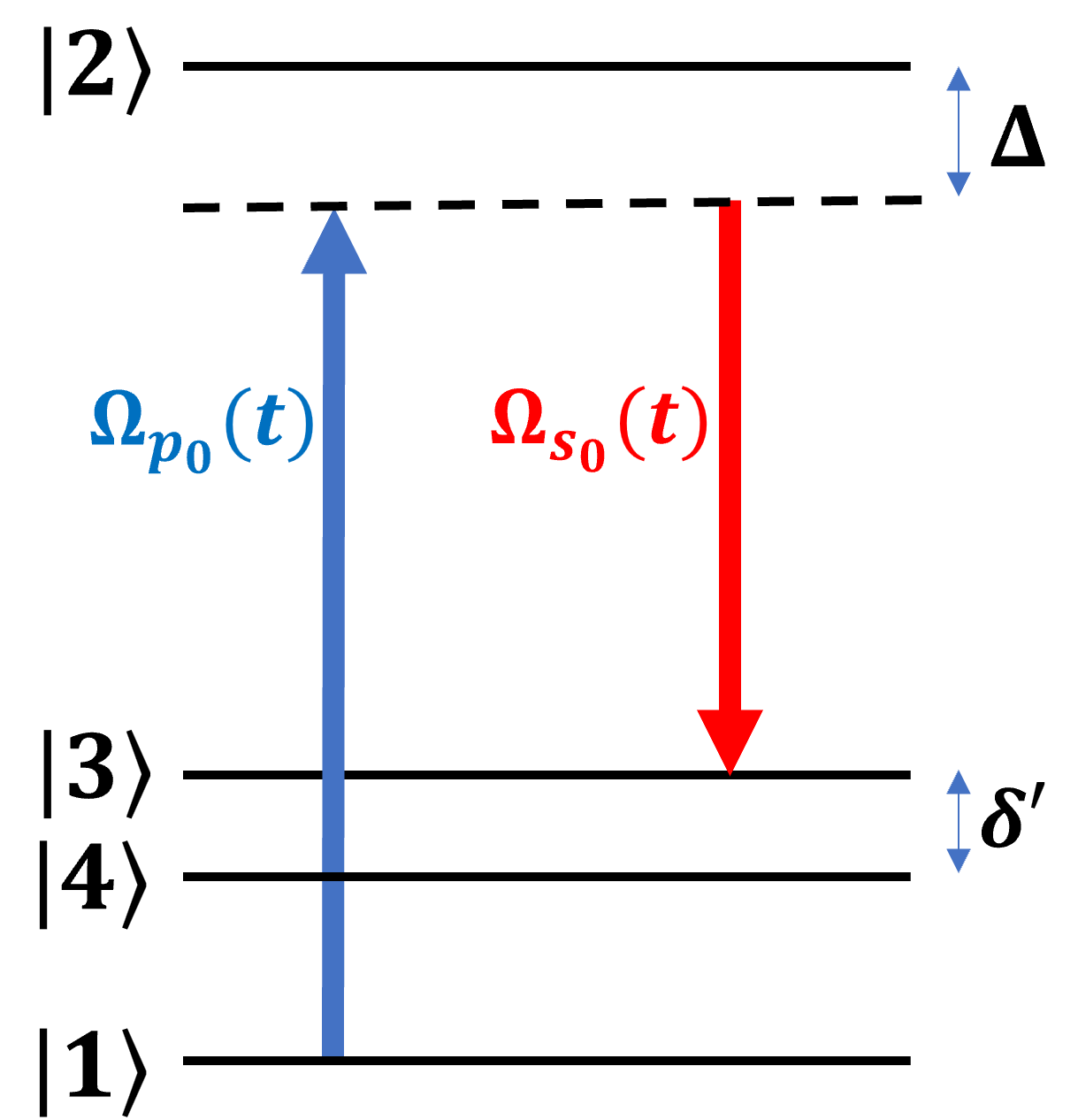}
\caption{The four-level $\lambda$ system of the STIRAP configuration having two energetically close final states. The two-photon resonance is with state $\ket{3}$ and the two-photon detuning from state $\ket{4}$ is $\delta'=\omega_p-\omega_s-(\omega_4-\omega_1)$; the one-photon detuning is $\Delta=\omega_2-\omega_1-\omega_p$.}\label{4-level}
\end{figure}

\begin{figure}    \includegraphics[scale=0.6]{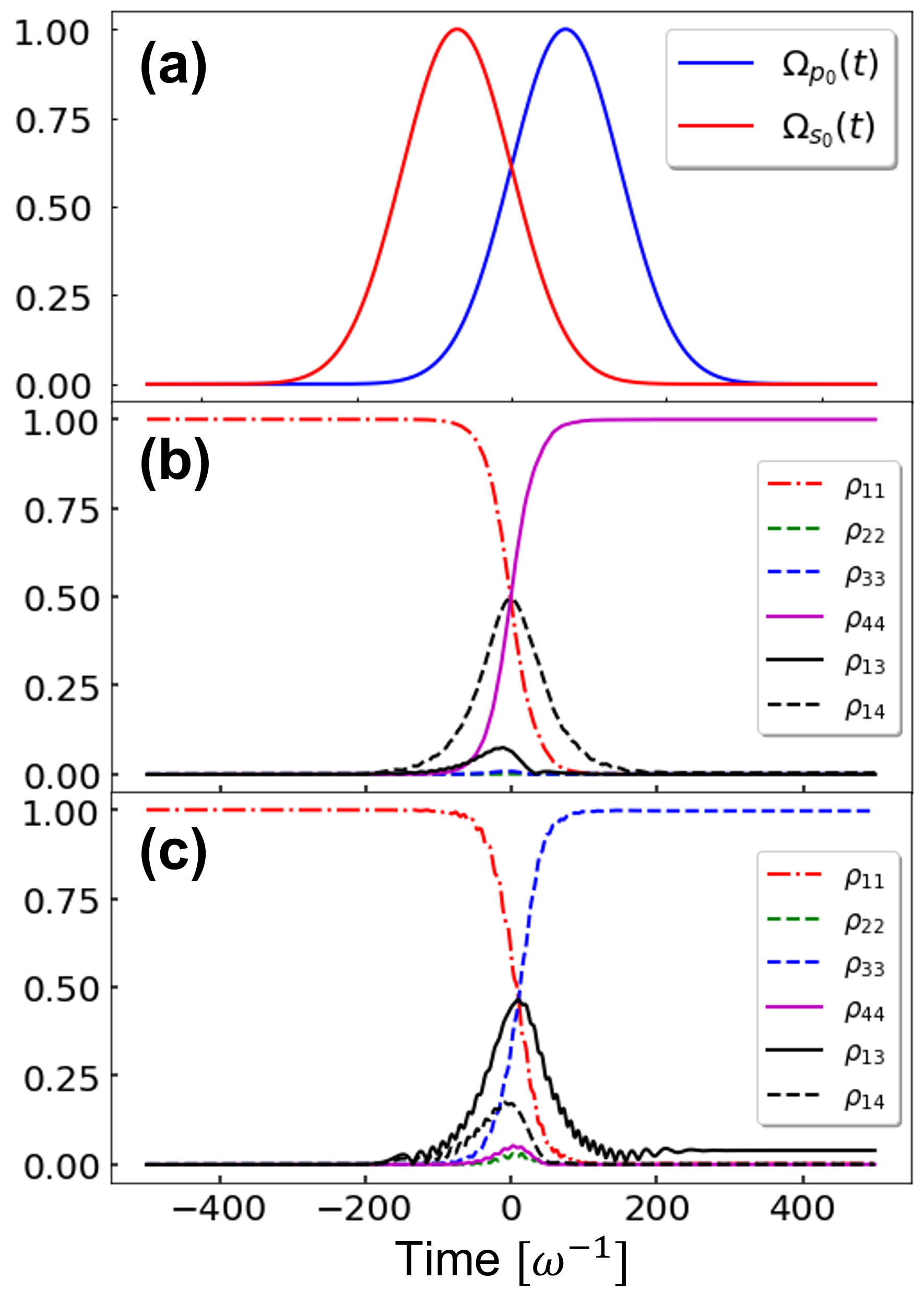}
\caption{The selective population transfer in the four-level STIRAP scheme achieved by controlling the sign of the chirp rate in the case of positive two-photon detuning, $\delta'=0.14[\omega]$. In (a), the Rabi frequencies are shown as a function of time; in (b), the detuned state $\ket{4}$ is populated owing to a positive chirp rate given by $\alpha=\delta'/\alpha(t_p-t_s)$ where $\alpha=\beta$. In contrast, in (c), the population is transferred to the resonant state $\ket{3}$ with a negative chirp rate given by $\alpha=-\delta'/\alpha(t_p-t_s)$. The parameters used are $\Delta=0$, $t_s=-70[\omega^{-1}]$, $t_p=70[\omega^{-1}]$, $ \tau_{p,s}=100[\omega^{-1}]$ and the peak Rabi frequencies are $\Omega_{{p,s}_0}=1.0[\omega]$. The values of the chirp rates are $\alpha = \pm 1\times10^{-3} [\omega^2]$.}\label{C-STIRAP_4-level_populations}
\end{figure}

Consider a system with an additional level nearly degenerate with the final state. 
A schematic of such a system is shown in Fig. \ref{4-level} with the two-photon resonance occuring with state $\ket{3}$, which implies  the two-photon detuning is  $\delta'=\omega_p-\omega_s-(\omega_4-\omega_1)$ and the one-photon detuning is $\Delta=\omega_2-\omega_1-\omega_p$. 
In the RWA and the field interaction representation, the Hamiltonian of the  four-level system is
\begin{equation}\label{FieldH4lvl}
H(t)=\frac{\hbar}{2}\left(\begin{array}{cccc}
0 & \Omega_{p_0}(t) & 0 & 0\\
\Omega_{p_0}(t) & 2\Delta(t) & \Omega_{s_0}(t) & \Omega_{s_0}(t)\\
0 & \Omega_{s_0}(t) & 2(\delta'(t)+\delta')& 0\\
0 & \Omega_{s_0}(t) & 0 & 2\delta'(t)
\end{array}\right), 
\end{equation}
where $\Delta(t)$ 
and $\delta'(t)$ are defined as  
$\Delta(t)=\Delta-\alpha(t-t_p)$ and $\delta'(t)= -\delta' + \beta(t-t_s)-\alpha(t-t_p)$. With the choice of equal chirp rates for the pump and the Stokes pulses, $\beta=\alpha$, the fourth diagonal term in the Hamiltonian can be cancelled out by fulfilling the condition $\alpha=\delta'/(t_p-t_s)$. This is a sufficient condition for transferring the population to the detuned state $\ket{4}$ adiabatically. This condition implies that both the chirp rate and the two-photon detuning $\delta'$ need to have the same sign. In contrast, choosing the chirp rate equal to  $\alpha=-\delta'/(t_p-t_s)$ 
results in transferring population to the resonant state $\ket{3}$; in this case, the signs of the detuning and the chirp have to be opposite. The dynamics of the selective population transfer to each  of the final states is shown in  Fig. \ref{C-STIRAP_4-level_populations}. 
In (a), the Rabi frequencies are shown as a function of time; in (b), the bare state $\ket{4}$ is populated at the end of pulse sequence with the  choice of $\alpha=1\times 10^{-3} [\omega^{2}]$; and in (c), the population is driven to the final bare state $\ket{3}$ with the choice of the negative chirp $\alpha=-1 \times 10^{-3} [\omega^{2}]$.

\begin{figure*}
\includegraphics[scale=0.6]{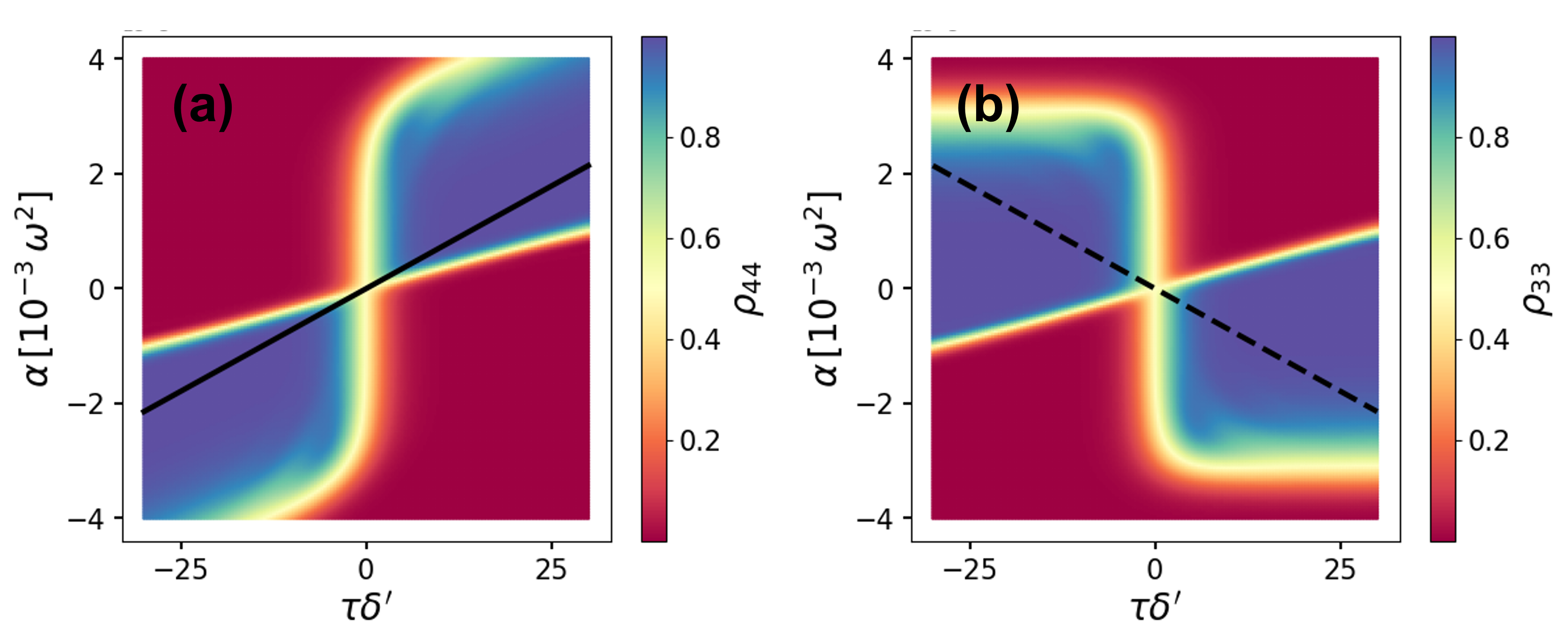}\caption{The populations of states $\ket{4}$ (a) and $\ket{3}$ (b) as a function of  $\delta'$ and  $\alpha$. The dark solid line represents the constraint  condition $\delta'(t)=-\delta'+\alpha(t_p-t_s)=0$ for the selective excitation of state $\ket{4}$.
For state $\ket{3}$ selective excitation, the constraint condition reads  $\delta'(t)=\delta'+\alpha(t_p-t_s)=0$,  
it is represented by the dashed line. 
}\label{populations_delta_vs_chirp_4-level}
\end{figure*}

Figure \ref{populations_delta_vs_chirp_4-level} shows the contour-plots of populations of states $\ket{4}$, (a), and $\ket{3}$, (b), as a function of detuning $\delta'$ and chirp rate $\alpha$. The dark solid line in (a) represents the constraint condition $-\delta'+\alpha(t_p-t_s)=0$. For selective excitation of state $\ket{4}$, the chirp and the detuning must be chosen in the vicinity of this line, implying that both of them have the same sign. The condition to drive the transition to state $\ket{3}$ is $\delta'+\alpha(t_p-t_s)=0$, it is represented by the dark dashed line in (b), implying that the signs of the detuning and the chirp need to be opposite. 
For zero detuning, states $\ket{3}$ and $\ket{4}$ are degenerate and are equally populated. When detuning deviates from zero, the detuned state is selectively populated if the chirp has the same sign as the detuning, or population goes solely to the resonant state if the signs of the chirp and the detuning are opposite. The latter case implies that the non-adiabatic term is not cancelled out as it was the case in the previous section.  
Such a condition implies the loss of adiabaticity characteristic for conventional STIRAP. The dressed state analysis indeed demonstrates that the evolution of the wavefuction in this case involves a series of dressed states. Surprisingly enough, even under the condition of non-adiabatic coupling between dressed states, the range of $\alpha$ and $\delta'$ parameters is broad and demonstrates the robustness of the approach.

Such a dependence of the state dynamics on the sign of detuning and the chirp is further explained in the next section by analysing the evolution of the dressed states in the field interaction frame.

\subsection{Dressed-state analysis of C-STIRAP in the four-level $\lambda$ system}

If in the three-level C-STIRAP the transformation to a dressed state basis can be done using a three dimensional rotation matrix, it is not so trivial in the case of the four-level C-STIRAP. For the adiabatic passage, the system has to remain in a single dressed state throughout its evolution and the non-adiabatic effects should be negligible. Here we present a numerical analysis of non-adiabaticity and population transfer based on the dressed state analysis in the field interaction picture and show that a scheme can be engineered in such a way that non-adiabatic effects are suppressed through proper choice of the detunings and the chirp rate.

The dressed state energies of the four-level system are the roots of the quartic polynomial $f(\lambda(t))$ obtained using the Hamiltonian in Eq. \eqref{FieldH4lvl}. The effect of the two fields is to shift the quartic equation, $f_0(\lambda(t))$, by a parabola, $f_1(\lambda(t))$.
The quartic polynomial $f(\lambda(t))$ reads 
\begin{align}\label{FieldH4lvl2}
\begin{split}
f(\lambda(t))&=f_0(\lambda(t))+f_1(\lambda(t))\\
f_0(\lambda(t))&=\left(\lambda(t)-2\delta'(t)\right)\left(\lambda(t)-2\delta'(t)+2\delta'\right)\lambda(t) \\ \nonumber
&\left(\lambda(t)-2\Delta(t)\right)\\
f_1(\lambda(t))&=2|\Omega_{s_0}(t)|^2\left(\lambda(t)(2\delta'(t)-\lambda(t))\right) \\ \nonumber
&-|\Omega_{p_0}(t)|^2\left(2\delta'(t)-\lambda(t)\right)\left(2\delta'(t)-2\delta'-\lambda(t)\right).
\end{split}
\end{align}
This introduces mixing of the in-going dressed states having energies $\lambda_k^{-}(t)$ (where $t\rightarrow-\infty$) with the out-going dressed states having energies $\lambda_k^{+}(t)$ (where $t\rightarrow+\infty$). 
The ingoing and outgoing dressed states  have energies given by the roots of $f_0(\lambda(t))$, which are obtained from  
the secular equation for the Hamiltonian \eqref{FieldH4lvl} in the limit of vanishing field strength. The $f_1(\lambda(t))$ is the remainder polynomial in the general case of the non-zero external fields. The in/out going dressed state energies are also the bare state energies.

An understanding of the non-adiabatic contributions requires a study of dynamics in the vicinity of avoided crossings between the dressed state energies, where $\left|\lambda_i(t)-\lambda_j(t)\right|/|\bra{\lambda_i(t)}\dot{H}(t)\ket{\lambda_j(t)}|\le 1$. 
The non-trivial coupling rate between two dressed states $\ket{\lambda_i}$ and $\ket{\lambda_j}$ is given by 
\begin{eqnarray}\label{nonad_coupling}
V_{ij}(t)&=&\abs{\braket{\lambda_i(t)}{\dfrac{d}{dt}\lambda_j(t)}} \\ \nonumber &=&\abs{\dfrac{\bra{\lambda_i(t)}\dot{H}(t)\ket{\lambda_j(t)}}{\lambda_i(t)-\lambda_j(t)}}\,, i \neq j \,.
\end{eqnarray}

\begin{figure*}
\includegraphics[scale=0.4]{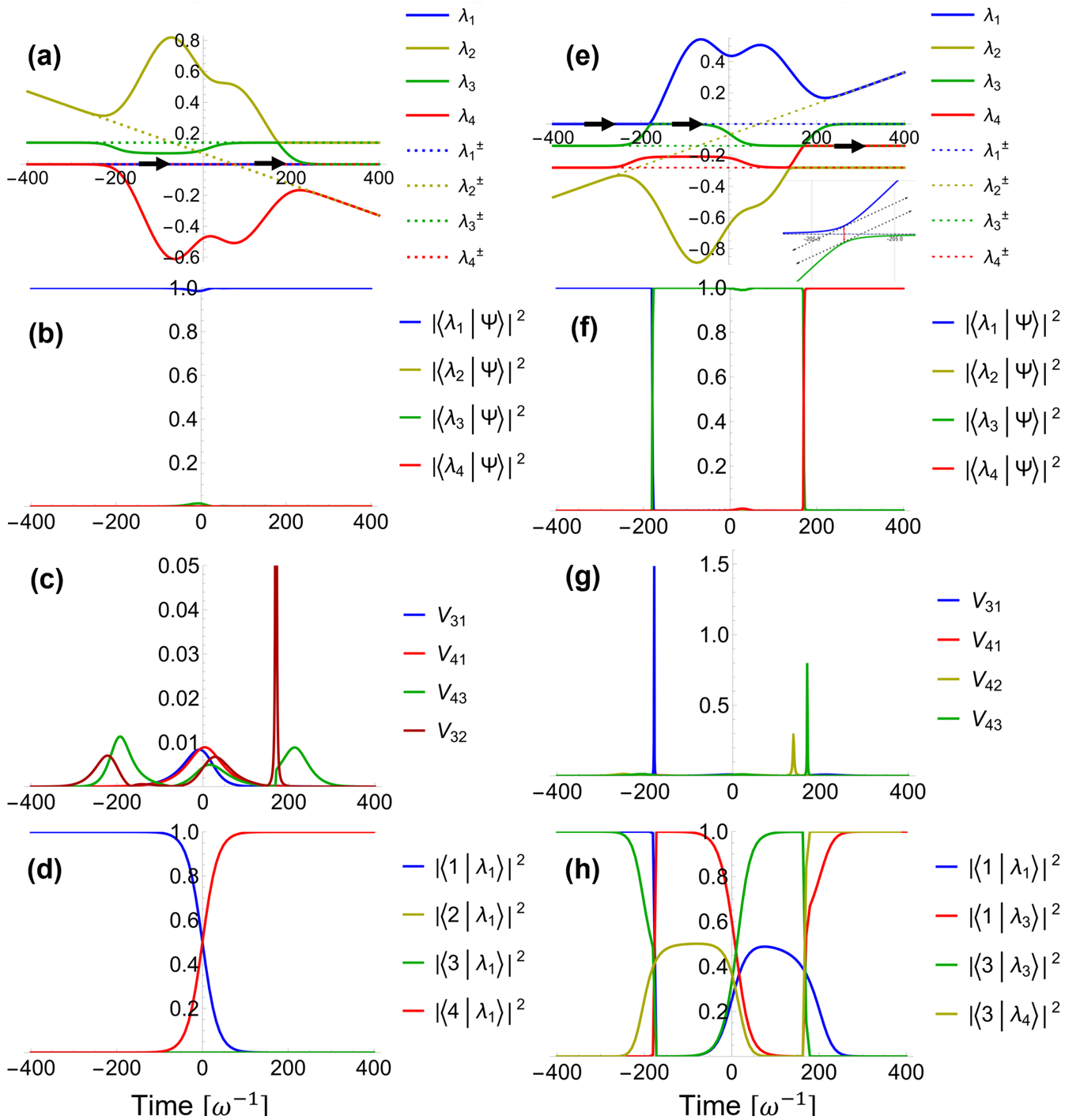}\caption{Dressed state analysis of the selective population transfer. Figures on the left, (a), (b), (c) and (d) correspond to the case with the positive chirp rate, dynamics for which is shown in Fig. \ref{C-STIRAP_4-level_populations}(b), and figures on the right, (e), (f), (g) and (h) correspond to the case with the negative chirp rate, related to Fig \ref{C-STIRAP_4-level_populations}(c).
The arrows in (a) and (e) represent the corresponding dressed state the system is aligned with during the evolution. The parameters used are $\Delta=0$, $t_s=-70[\omega^{-1}]$, $t_p=70[\omega^{-1}]$, $ \tau_{p,s}=100[\omega^{-1}]$ and the peak Rabi frequencies are $\Omega_{{p,s}_0}=1.0[\omega]$. The chirp rate is $\alpha = \pm1\times10^{-3} [\omega^2]$ for the left and right figures respectively. 
}\label{4-level_C-STIRAP_dressed_states}
\end{figure*}

The numerical dressed state analysis of the selective population transfer in the four-level system 
is presented in Fig. \ref{4-level_C-STIRAP_dressed_states}, where the left figures, Figs.\ref{4-level_C-STIRAP_dressed_states}(a), \ref{4-level_C-STIRAP_dressed_states}(b), \ref{4-level_C-STIRAP_dressed_states}(c) and \ref{4-level_C-STIRAP_dressed_states}(d), correspond to the case of the positive chirp rate shown in Fig. \ref{C-STIRAP_4-level_populations}(b), and the right figures, Figs.\ref{4-level_C-STIRAP_dressed_states}(e), \ref{4-level_C-STIRAP_dressed_states}(f), \ref{4-level_C-STIRAP_dressed_states}(g) and \ref{4-level_C-STIRAP_dressed_states}(h), correspond to the case of the negative chirp rate shown in Fig. \ref{C-STIRAP_4-level_populations}(c). In \ref{4-level_C-STIRAP_dressed_states}(a) and \ref{4-level_C-STIRAP_dressed_states}(e), the bare state energies, $\lambda_k^{\pm}(t)$ (in/out going dressed energies), are represented by dashed lines and the dressed state energies are represented by solid lines. A single dressed state $\lambda_k(t)$ does not connect to the same in-going and out-going dressed states $\lambda_k^{\pm}(t)$. While the two $\lambda_k^{\pm}(t)$ are represented by a single dashed line, the time-evolution of the k-th dressed state $\lambda_k(t)$ results in its connection to the different outgoing $\lambda_m^{+}(t)$ state. 
At initial time $t\rightarrow{}-\infty$, the energy of  eigenstate $\ket{\lambda_1(t)}$ starts in zero energy of the ground bare state $\ket{1}$. In the case of positive chirp rate, the mechanism of selective population of the bare state $\ket{4}$ is through providing resonance 
via compensating the positive value of two-photon detuning by the choice of the positive chirp, such that $\delta'(t)=-\delta' + \alpha(t_p-t_s)=0$. This causes $\lambda_1(t)\approx0$ for all time, as seen in \ref{4-level_C-STIRAP_dressed_states}(a), and results in a degeneracy for the bare states $\lambda_{1}^{\pm}(t)=\lambda_{4}^{\pm}(t)$. Owing to this, during the evolution, the system stays in a  single dressed state $\ket{\lambda_1(t)}$, which evolves from bare state $\ket{1}$ to $\ket{4}$, \ref{4-level_C-STIRAP_dressed_states}(d). The solid arrows along $\lambda_1(t)$ indicate that the system is in the respective dressed state $\ket{\lambda_1(t)}$ all the time. The dressed state $\ket{\lambda_1(t)}$ is the dark state. 
The probability amplitudes of the dressed states with respect to the statevector $\ket{\psi(t)}$, shown in \ref{4-level_C-STIRAP_dressed_states}(b), confirm the adiabatic evolution of the wave function along the dressed state $\ket{\lambda_1(t)}$,  since it is isolated from the rest manifold. The non-adiabatic coupling terms, shown in \ref{4-level_C-STIRAP_dressed_states}(c), with an exception of the coupling between $\ket{\lambda_2(t)}$ and $\ket{\lambda_3(t)}$, are an order of magnitude less than the closest separation of the dressed states, which confirms the adiabatic passage as a mechanism of population transfer. The high value of coupling rate $\braket{\lambda_3(t)}{\tfrac{d}{dt}\lambda_2(t)}$, which coincides with the avoided crossing between the corresponding eigenenergies at $t\approx 200[\omega^{-1}] $, does not adversely affect the adiabaticity as the total wavefunction aligns with $\ket{\lambda_1(t)}$ at all times.





In contrast to the previous case, the time evolution of dressed states for 
negative value of the chirp, shown in Fig. \ref{4-level_C-STIRAP_dressed_states}(e)-(h), demonstrates a complete population transfer  from $\ket{1}$ to the resonant state  $\ket{3}$  
via a non-adiabatic process involving three dressed states. Two non-adiabatic transitions occur at a low field intensity while a smooth, an adiabatic type, population transfer takes place within a single dressed state at strong fields. 
In more detail, at time $ t\approx -200 [\omega^{-1}]$,  originally populated dressed state $\ket{\lambda_1(t)}$, keeping population within the bare state $\ket{1}$, approaches an avoided crossing with $\ket{\lambda_3(t)}$, shown in Fig.\ref{4-level_C-STIRAP_dressed_states}(e), and transfers population there owing to non-zero coupling shown in Fig.\ref{4-level_C-STIRAP_dressed_states}(g). 
Further time evolution takes place within $\ket{\lambda_3(t)}$ during which bare states $\ket{1}$ and $\ket{3}$ exchange adiabatically population. At time $ t\approx 200 [\omega^{-1}]$, the second avoided crossing occurs of $\ket{\lambda_3(t)}$ with dressed state $\ket{\lambda_4(t)}$.  Here the population is transferred again populating bare state $\ket{3}$ within $\ket{\lambda_4(t)}$ by the end of pulses' duration, as shown in Fig.\ref{4-level_C-STIRAP_dressed_states}(h). The non-adiabatic couplings, shown in Fig.\ref{4-level_C-STIRAP_dressed_states}(g), at the same times as the avoided crossings,  provide population transfer  between respective dressed states. The slopes of the dressed state energies curves  and the closest approach distance between the curves at the crossing give us an estimate of the transition probability. The closest approach distance is $8 \times 10^{-4}$ $[\omega]$ at $t_a=198.1$ $[\omega^{-1}]$, and $\left|\dfrac{d}{dt}(\lambda_3(t_a)-\lambda_4(t_a))\right|=3.48 \times 10^{-4}$ $[\omega]$, and the non adiabatic coupling, $\braket{\lambda_4(t)}{\dfrac{d}{dt}\lambda_3(t)}$, is a Lorentzian curve centered at $t=t_a$ with width $w=0.35$ and area $A=\pi/2$. 
The population is transferred from bare state $\ket{1}$ to bare state $\ket{3}$ non-adiabatically, owing to synergistic dynamics between dresses states shown in Fig.\ref{4-level_C-STIRAP_dressed_states}(f), first $\ket{\lambda_1}(t)$ and $\ket{\lambda_3}(t)$, and then $\ket{\lambda_3}(t)$ and $\ket{\lambda_4}(t)$. Notably, the majority of population transfer 
occurs during the time the two pulses overlap.

\begin{figure*}
\includegraphics[scale=0.8]{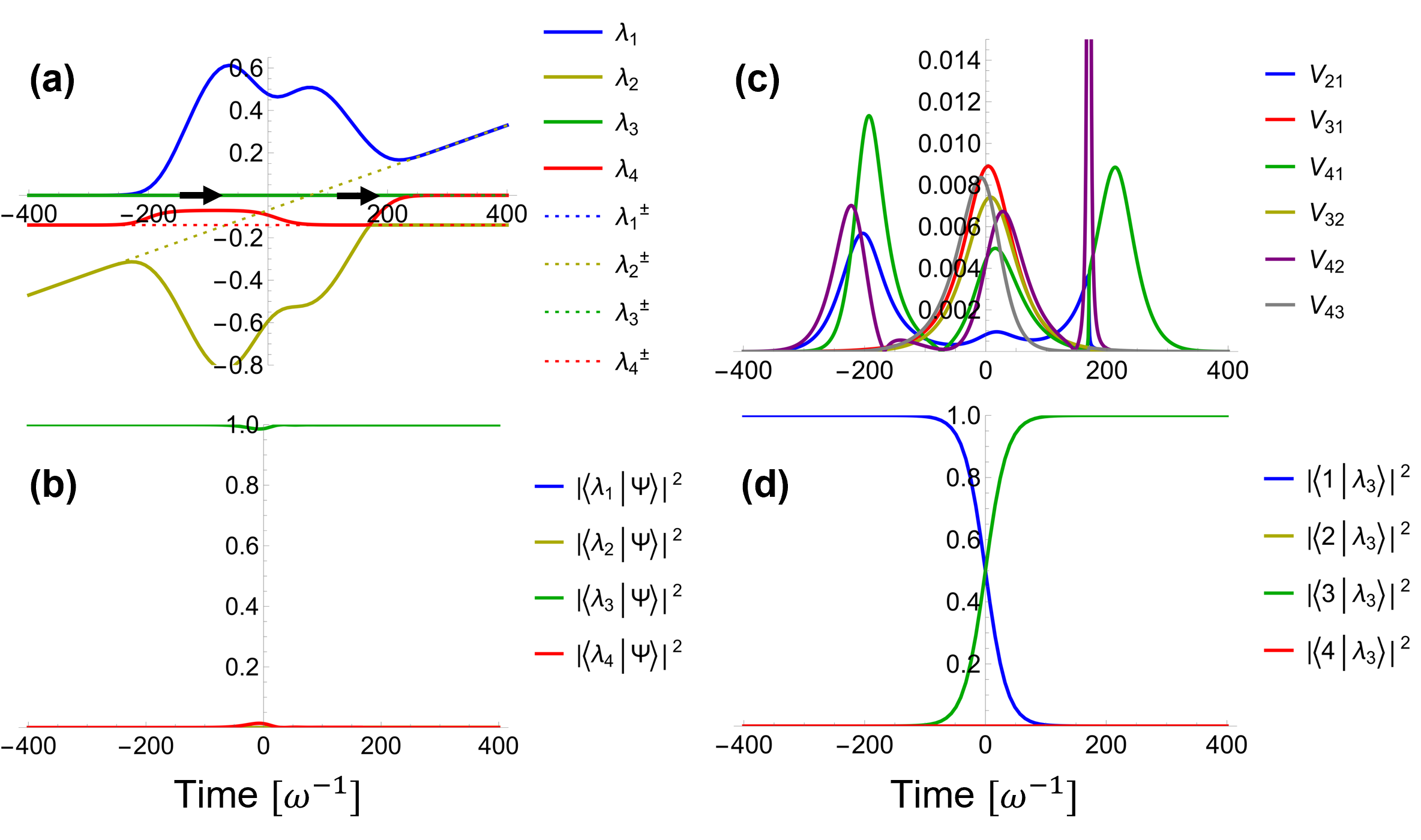}\caption{Adiabatic population transfer to the resonant state $\ket{3}$ by applying a chirping delay $t_d=t_p-t_s$ in the Stokes pulse. The wavefunction $\ket{\psi(t)}$ is always aligned with the dressed state $\ket{\lambda_{3}(t)}$, (b), which smoothly evolves from bare state $\ket{1}$ to $\ket{3}$, (d). There is an avoided between states $\ket{2}$ and $\ket{4}$ implying a high value of $V_{42}=\braket{\lambda_4(t)}{\tfrac{d}{dt}\lambda_2(t)}$. The coupling does not include the dark state $\ket{3}$ confirming that the process is adiabatic.
}\label{4-level_C-STIRAP_rho33_dressed_states}
\end{figure*}

\begin{figure}
\includegraphics[scale=0.6]{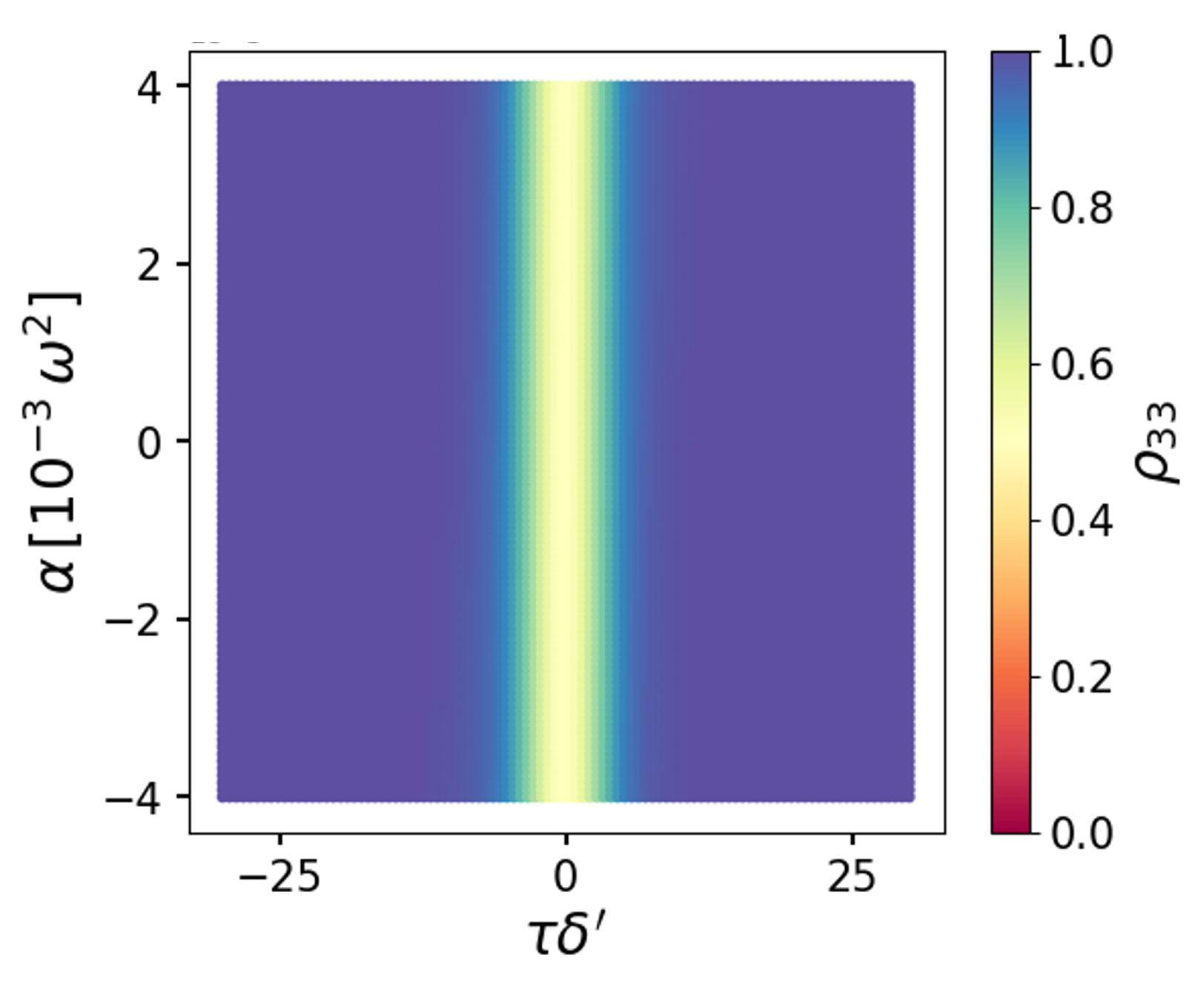}\caption{The population of resonant state $\ket{3}$ as a function of  $\tau*\delta'$ and chirp rate $\alpha$ for the case when a chirping delay $t_d = t_p - t_s$ is applied in the Stokes pulse. A full population transfer to the state $\ket{3}$ occurs adiabatically for all chirp values in this range owing to the degeneracy between states $\ket{1}$ and $\ket{3}$. }\label{4-level_C-STIRAP_rho33_contour}
\end{figure}

As demonstrated in Fig. \ref{4-level_C-STIRAP_dressed_states}, the population transfer to the resonant state $\ket{3}$ is not an adiabatic process. However, it is possible to transfer the population to $\ket{3}$ via adiabatic passage by introducing a chirping delay in the Stokes pulse. The modified Stokes pulse with delay $t_d$ reads:
\begin{equation}\label{C-STIRAP_delayed_pulse}
E_{s}(t)=E_{s_0}  e^{\frac{-(t-t_{s})^2}{\tau_{s}^2}} \cos[\omega_{s}(t-t_{s})+\tfrac{1}{2}\beta(t-t_{s}-t_d)^2].
\end{equation}
This modifies the $\delta'(t)$ in Eq. \eqref{FieldH4lvl} to $\delta'(t)=-\delta' +\beta(t-t_s-t_d)-\alpha(t-t_p)$. With a choice of $t_d=t_p-t_s$, the third diagonal element is cancels out making the states $\ket{1}$ and $\ket{3}$ degenerate. This is the condition to populate $\ket{3}$ adiabatically. The evolution of dressed state energies in this case is given in Fig. \ref{4-level_C-STIRAP_rho33_dressed_states}(a). Here, the dark state is $\ket{\lambda_{3}(t)}$, which the system is always aligned with, as seen in Fig. \ref{4-level_C-STIRAP_rho33_dressed_states}(b), smoothly evolves from bare state $\ket{1}$ to bare state $\ket{3}$, shown in Fig. \ref{4-level_C-STIRAP_rho33_dressed_states}(d). All the non-adiabatic coupling rates are negligible compared to the dressed state energy separations confirming the passage is adiabatic. In Fig. \ref{4-level_C-STIRAP_rho33_contour}, a contour plot of population $\rho_{33}$ is depicted as function of the two-photon detuning $\delta'$  and the chirp rate $\alpha$. The figure demonstrates that, for adiabatic population transfer to state $\ket{3}$, there is no constraint condition on the value of the chirp rate within the given range as long as a delay $t_d$ is applied and satisfies the above condition. 

The selectivity of the final state excitation is possible only for small values of chirp rates satisfying the Landau-Zener adiabaticity condition requiring $\Omega_{p_0,s_0}^2/\alpha \gg 1$. For larger values of chirp rates, adiabatic passage is not possible leading to an arbitrary superposition of states $\ket{3}$ and $\ket{4}$.


\section{Chirped Fractional STIRAP (C-F-STIRAP)}

\subsection{F-STIRAP}

\begin{figure}	\includegraphics[scale=0.5]{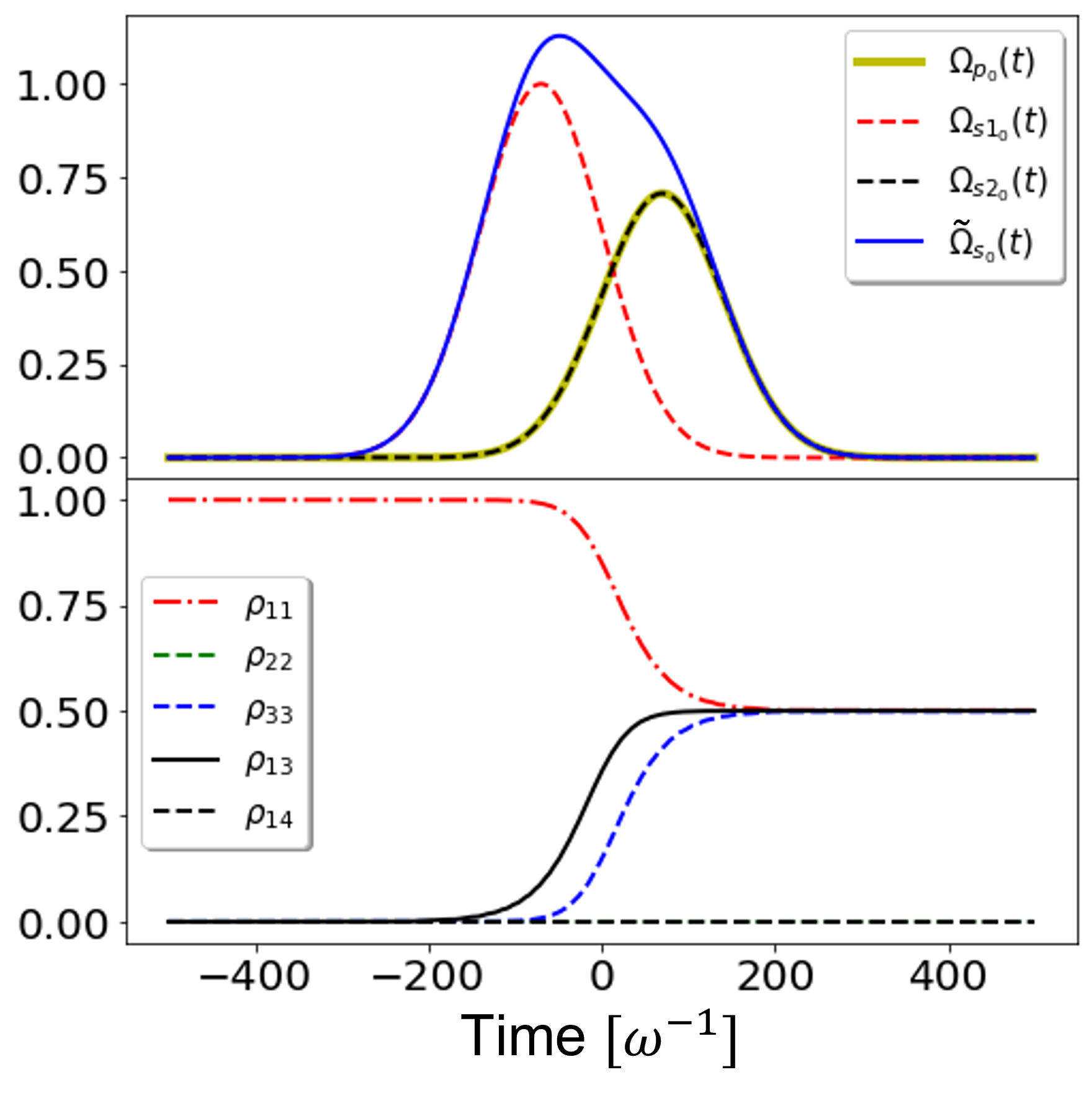}	\caption{F-STIRAP obtained via the superposition of two Stokes pulse components. 
In (a), the resultant Stokes field is represented by the solid line. Here the constant mixing angle is $A=\pi/4$. Note that the pump pulse $\Omega_{p_0}(t)$ overlaps exactly with the second Stokes pulse component $\Omega_{s2_0}(t)$. In (b), populations and coherence are shown as a function of time.}\label{F-STIRAP}
\end{figure}

The notion of fractional STIRAP, F-STIRAP, is to partially preserve the population in the ground state, thus, creating a coherent superposition of the initial and the final states by the end of the pulse sequence. 
The approach implies elongating the 
Stokes pulse so that it would vanish simultaneously with the pump pulse. This requires that the value of mixing angle $\theta(t)=\tan^{-1}\left[\Omega_{p_0}(t)/\Omega_{s_0}(t)\right]$, which for conventional STIRAP is $\theta(t\rightarrow -\infty)=0$ and $\theta(t\rightarrow \infty)=\pi/2$, while for F-STIRAP it is less than $\pi/2$ as $t\rightarrow\infty$. It is possible to control the asymptotic value of the mixing angle by modifying the pump and the Stokes fields as follows

\begin{equation}
\begin{aligned}
E_{p}(t) &= E_{p_0} \sin A e^{-\tfrac{(t-t_p)^2}{\tau^2}}\cos[\omega_{p}(t-t_{p})]
\\
E_{s}(t) &= E_{s_0} e^{-\tfrac{(t+t_p)^2}{\tau^2}}\cos[\omega_{s}(t+t_{p})] \\ \nonumber &+ E_{s_0} \cos A e^{-\tfrac{(t-t_p)^2}{\tau^2}}\cos[\omega_{s}(t-t_{p})]\,,  \\
\end{aligned}
\label{F-STIRAP_fields}
\end{equation}
where the Stokes field  is composed of two Stokes pulses each having central time $t_p$ and $-t_p$, with $t_p$ being the central time of the pump pulse as well. The
angle $A$ is the constant mixing angle, here 
equivalent to $\theta(t\rightarrow\infty$). Substituting these fields in the Hamiltonian in Eq. \eqref{Hamil-Schrodinger}, and applying the transformations $a_1 = \tilde{a_1} e^{i\omega_p(t-t_p)}, a_2 = \tilde{a_2}$ and $a_3 = \tilde{a_3} e^{i\omega_s(t+t_p)}$ in the Schr\"odinger Eq.\eqref{Schrodinger} provide the field-interaction Hamiltonian of the three-level F-STIRAP scheme:

\begin{gather}
\scalebox{0.95}{$
\begin{aligned}
&\mathbf{H}(t) = \frac{\hbar}{2} \times \\ 
& \left( \begin{array}{cccc} 0   & \Omega_{p_0} (t)    &   0\\
	\Omega_{p_0}(t) &   2\Delta &   \Omega_{s1_0}(t)+\Omega_{s2_0}(t)e^{i\phi} \\
	0   &   \Omega_{s1_0}(t)+\Omega_{s2_0}(t)e^{-i\phi}   &  -2\delta\\
\end{array} \right), 
\end{aligned}$}
\label{Ham3level-F-STIRAP}
\end{gather}
where the Rabi frequencies are
\begin{equation}\label{F-STIRAP_envelopes}
\begin{aligned}
\Omega_{p0}(t) &= \Omega_{0} \sin A e^{-\tfrac{(t-t_p)^2}{\tau^2}} \\
\Omega_{s10}(t) &=  \Omega_{0} e^{-\tfrac{(t+t_p)^2}{\tau^2}}\\
\Omega_{s20}(t) &= \Omega_0 \cos A e^{-\tfrac{(t-t_p)^2}{\tau^2}}\,, \\
\end{aligned}
\end{equation}
with $\Omega_0=-E_{p0}\mu_{21}/\hbar=-E_{s0}\mu_{32}/\hbar$ and the phase is $\phi=2\omega_s t_p$. 
With an appropriate choice of $t_p$ satisfying the condition, $2\omega_st_p=2n\pi, n=1,2,3..$, $t_p=n\pi/\omega_s$ the phase dependence is cancelled out, e.g., for $t_p=71, \omega_s=5, $ $e^{-i\phi}\approx1$.
Note that for $A=\pi/2$, the second Stokes component is zero and the conventional STIRAP Hamiltonian is  retrieved. When $A=\pi/4$, the $\Omega_{p0}(t)$ perfectly overlaps with the second component of the Stokes pulse   $\Omega_{s20}(t)$, and they vanish simultaneously as shown in Fig. \ref{F-STIRAP}. Such an arrangement of pulses maximizes coherence between the initial and the final state. In general, the constant mixing angle  $A$ in the range of $\pi/4 \geq A \geq \pi/2$ creates an arbitrary superposition state having coherence up to its maximum value $1/2$. Throughout this section, the peak Rabi frequency and time duration are chosen to be $\Omega_0=1[\omega]$ and $\tau = 100 [\omega^{-1}]$.

\begin{figure}  \includegraphics[scale=0.48]{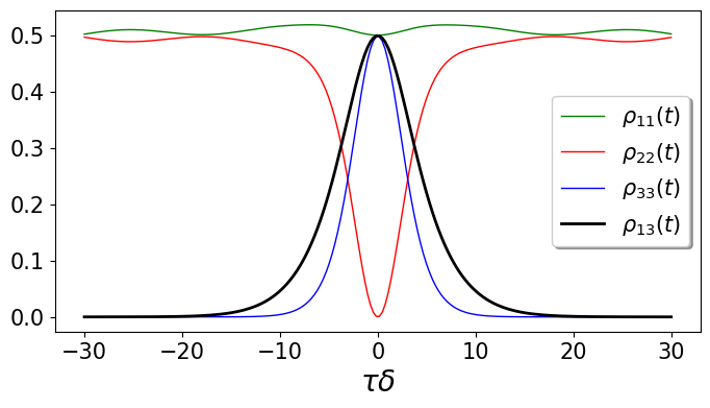}
\caption{ The dependence of coherence and populations on the two-photon detuning  in F-STIRAP.  The two-photon resonance is the necessary condition to create a maximally coherent superposition state.}\label{detuning_vs_coherence}
\end{figure}

As it was shown in the previous section, in the absence of pulse chirping, the two-photon resonance is required to achieve a complete adiabatic population transfer in STIRAP. Analogously, the two-photon resonance is required to maximize the coherence in the F-STIRAP in the absence of chirping. We investigated how coherence reduces from the maximum value with the increase of the two-photon detuning in the F-STIRAP configuration. Coherence between the initial and the final states as well as the population of these states as a function of  the two-photon detuning are shown in Fig. \ref{detuning_vs_coherence}. The reduction of coherence is observed from the maximum value by the factor of 2 for $\tau \delta=5$.

\subsection{F-STIRAP using a single, shaped Gaussian  Stokes pulse}

\begin{figure}	\includegraphics[scale=0.55]{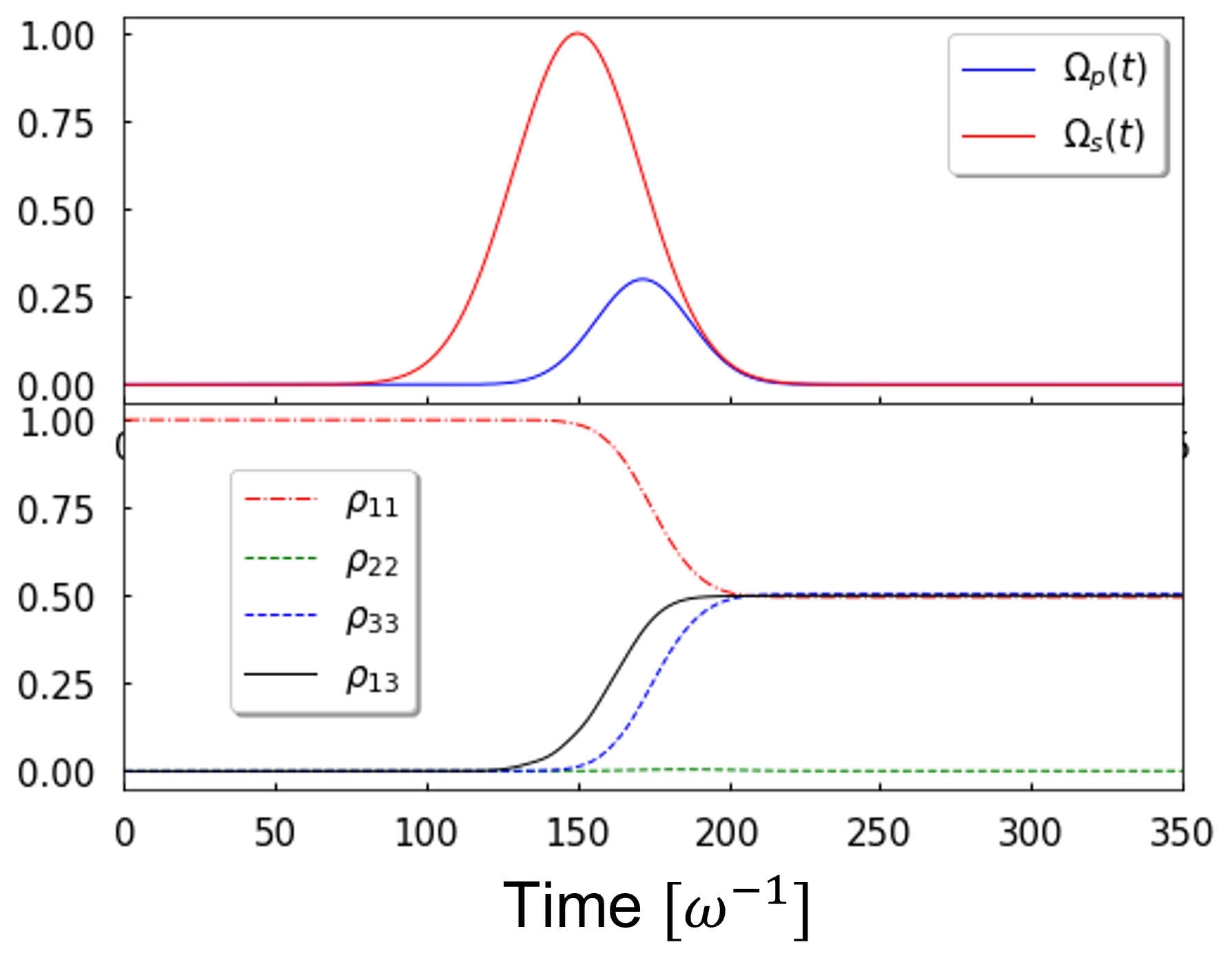}		\caption{F-STIRAP performed using a single Gaussian Stokes pulse. The maximum coherence between the initial and the final states is achieved by making the two pulses overlap before they vanish. The parameters are $t_p-t_s=21.6 [\omega^{-1}]$, $\Omega_p\tau_p=6.6$ and $\Omega_s\tau_s=30$.
}\label{G-F-STIRAP}
\end{figure}

\begin{figure} \includegraphics[scale=0.6]{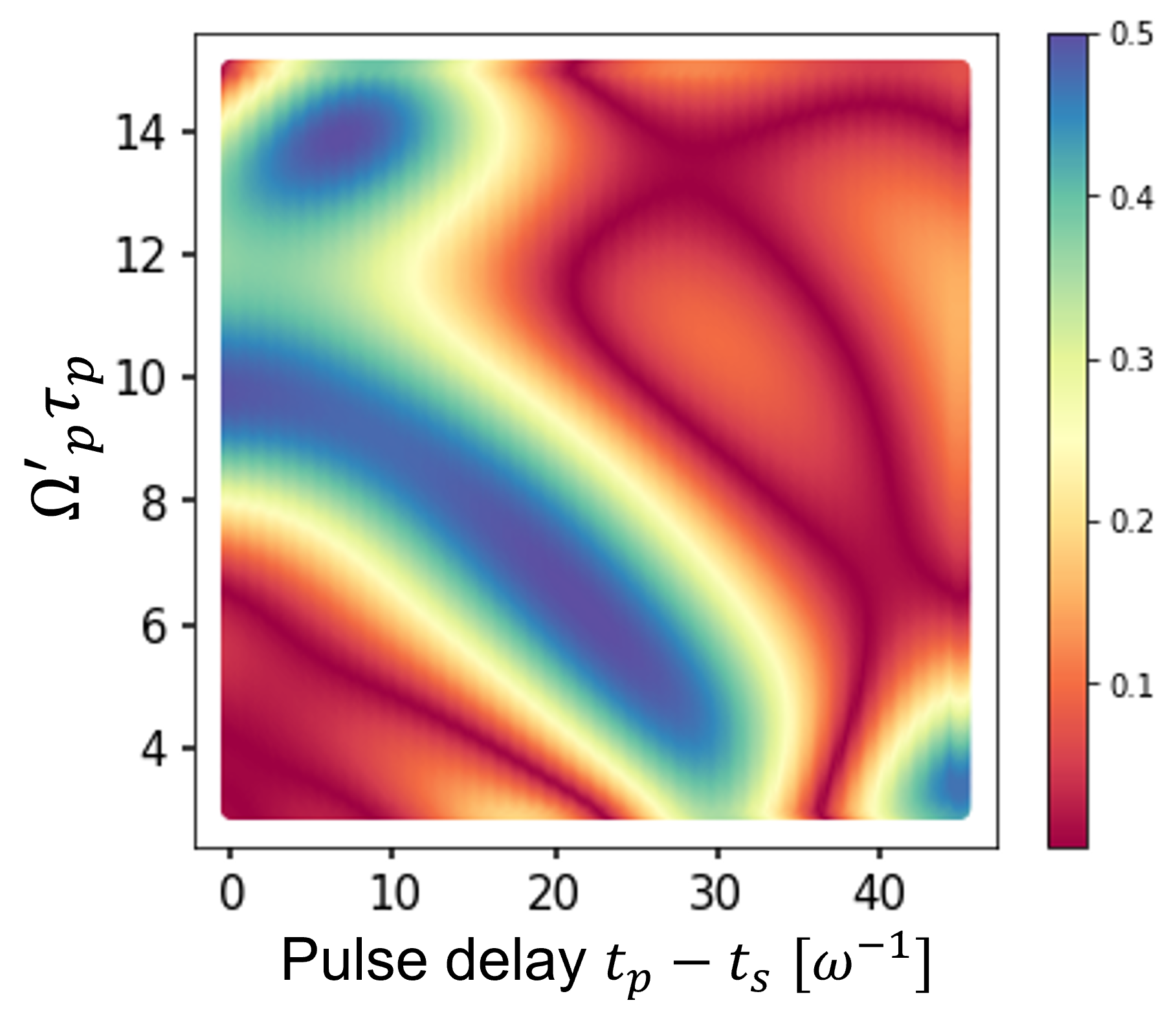}
\caption{Coherence $\rho_{31}$ between the initial and the final states as a function of the $\Omega_p \tau_p$ and pulse delay $(t_p-t_s)$ in the F-STIRAP using one Gaussian Stokes pulse. The pulse area and the central time of the Stokes pulse are kept constant while those of the pump pulse are varied. The parameters are $\Omega_s = 1[\omega] $, $\Omega_p= 0.3[\omega] $, $\tau_s = 30.0[\omega^{-1}]$, $t_s=150.0[\omega^{-1}] $. 
}\label{G-F-STIRAP_Contour}
\end{figure}

Creating a maximally coherent superposition of the initial and the final states is possible without the second component of the Stokes pulse. This can be done by carefully choosing the pulse parameters in such a way that the single Stokes pulse and the pump pulse overlap significantly before they vanish. Such a choice of the pump and the Stokes pulses, whose Rabi envelopes are given by $\Omega_{p,s}(t)=\Omega_{p,s}  \exp{-(t-t_{p,s})^2/\tau_{p,s}^2}$, and the population dynamics are given in Fig. \ref{G-F-STIRAP}. Here, the pulse delay is $t_p-t_s=21.6 [\omega^{-1}]$ and the pulse area of the pump and Stokes pulses  are $\Omega_p\tau_p=6.6$ and $\Omega_s\tau_s=30$. To demonstrate the robustness in this process, the coherence is plotted in Fig. \ref{G-F-STIRAP_Contour} as a function of the pulse delay and the pulse area of the pump pulse, $\Omega_p\tau_p$, keeping the parameters of the Stokes pulse constant at $\Omega_s = 1[\omega] $, $\tau_s = 30.0[\omega^{-1}]$, $t_s=150.0[\omega^{-1}] $. On the y-axis, the Rabi frequency of the pump pulse is kept constant while varying the time duration $\tau_p$. The area of the maximum coherence is observed in blue color for a broad range of the pump pulse area and the pulse delay up to the value 30 $[\omega^{-1}]$.


\subsection{C-F-STIRAP in the three-level $\lambda$ system} 
As demonstrated in Section III A, the two-photon resonance is required for the maximum coherence in F-STIRAP. Here, we introduce the chirped F-STIRAP (C-F-STIRAP) as the means to maximize coherence when the two-photon detuning is nonzero.
Consider the pump, the first and the second  Stokes pulse components are chirped with chirp rates $\alpha$, $\beta_1$ and $\beta_2$ respectively

\begin{gather}\label{F-STIRAP_envelopes_CHRPT}
\scalebox{0.88}{$\begin{aligned}
E_{p}(t) &= E_{p_0} \sin A e^{-\tfrac{(t-t_p)^2}{\tau^2}}\cos[\omega_{p}(t-t_{p})+\tfrac{1}{2}\alpha(t-t_{p})^2]
\\
E_{s}(t) &= E_{s_0} e^{-\tfrac{(t+t_p)^2}{\tau^2}}\cos[\omega_{s}(t+t_{p})+\tfrac{\beta_1}{2}(t+t_{p}-t_{d1})^2] \\
&+ E_{s_0} \cos A e^{-\tfrac{(t-t_p)^2}{\tau^2}}\cos[\omega_{s}(t-t_{p})+\tfrac{\beta_2}{2}(t-t_{p}-t_{d2})^2],  \\
\end{aligned}$}
\end{gather}
where the chirping of the first and the second Stokes pulses are assumed to have delays $t_{d1}$ and $t_{d2}$ respectively.
To derive the field-interaction Hamiltonian  the following transformations are applied 

\begin{figure*}
\includegraphics[scale=0.55]{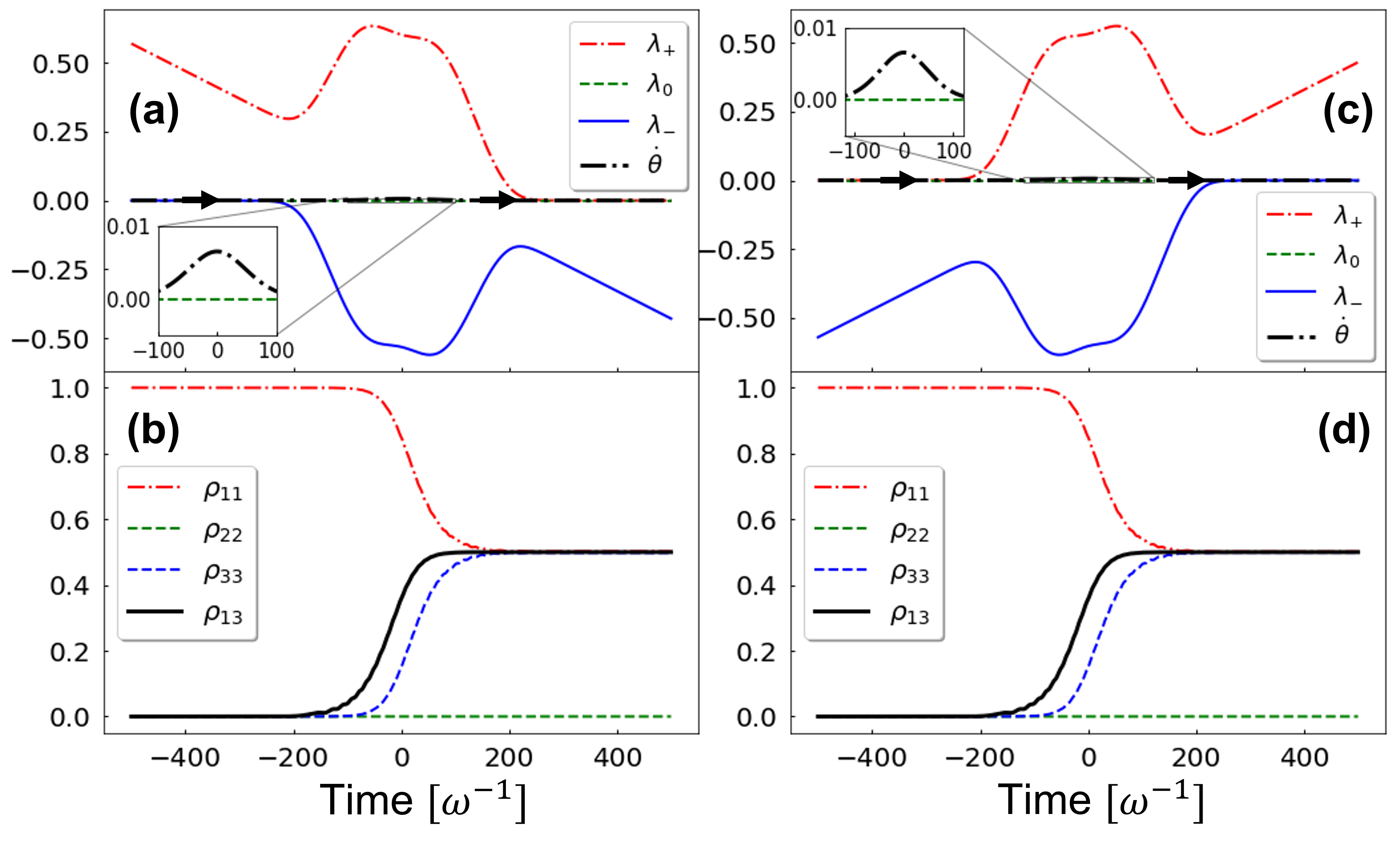}
\caption{The evolution of the Rabi frequencies, the  dressed states, the non-adiabatic parameter $V_{ij}$ and the state  coherence in the case of C-F-STIRAP for the detuning $\delta > 0$  (left) and $\delta < 0$ (right). A maximum coherence is created in both cases because of the choice of chip rates satisfying $\delta(t)=0$. The value of $V_{ij}$ remains zero except for a small duration as shown in the inset. The system remains in the dark state $\ket{\lambda_0}$ throughout the evolution, as indicated by the arrows. In (a) and (b), $\delta=-0.14 [\omega]$, $\alpha = -1\times10^{-3} [\omega^2]$ and in (c) and (d), $\delta=0.14 [\omega]$, $\alpha = 1\times10^{-3} [\omega^2]$. Other parameters are $\Delta=0$, $ \tau_{p,s}=100[\omega^{-1}]$ and peak $\Omega_{{p0,s0}}=1.0[\omega]$.}\label{C-F-STIRAP_dressed}
\end{figure*}

\begin{equation}\label{transf_chirped}		\begin{aligned}
a_1(t) &=\tilde{a_1}(t) e^{i\omega_p(t-t_p)} \\
a_2 (t)&= \tilde{a_2}(t) e^{-i\tfrac{\alpha}{2}(t-t_p)^2} \\
a_3(t) &= \tilde{a_3}(t) e^{i\omega_s(t+t_p) + i\frac{\beta_1}{2}(t+t_p-t_{d1})^2-i\frac{\alpha}{2}(t-t_p)^2}.
\end{aligned}
\end{equation}
Then the Hamiltonian describing C-F-STIRAP in the field-interaction representation reads
\begin{gather}\label{Ham3level_F-STIRAP}
\scalebox{0.9}{$\begin{aligned}
&\mathbf{H}(t) = \frac{\hbar}{2} \times \\
&\left( \begin{array}{ccc} 0   & \Omega_{p0} (t)    &   0	 \\
	\Omega_{p0} (t) &  2\Delta(t) &  \Omega_{s10}(t)+\Omega_{s20}(t)e^{i\eta(t)} \\
	0  &  \Omega_{s10}(t)+\Omega_{s20}(t)e^{-i\eta(t)} &  2\delta(t)\\
\end{array} \right),
\end{aligned}$}
\end{gather}
where the Rabi frequencies are the same as in Eq. \eqref{F-STIRAP_envelopes}, the time dependent detuning $\delta(t)=-\delta+\beta_1(t+t_p-t_{d1})-\alpha(t-t_p)$ and the time-dependent phase $\eta(t)$ is given by
\begin{gather}
\scalebox{0.9}{$\eta(t)=2\omega_st_p+\frac{\beta_1}{2}(t+t_p-t_{d1})^2-\frac{\beta_2}{2}(t-t_p-t_{d2})^2.$}
\end{gather}

If all the chirp rates are equal, $\alpha=\beta_1=\beta_2$, and the chirping delays are chosen to be $t_{d1}=0$, 
then the time dependent detuning $\delta(t)$ casts to $\delta(t) = -\delta + 2\alpha t_p$.
Besides, if the delay $t_{d2}$ is given by the negative time difference between the peaks of two Stokes pulses, $t_{d2}=-2t_p$,
then the phase $\eta(t)$ becomes a constant independent from the chirp rates, $\eta(t) = 2\omega_s t_p=\phi$. These assumptions result in the real values of the Rabi frequencies and the straightforward condition for the resonance with state $\ket{3}$ $\alpha=\delta/2t_p$.

The Hamiltonian in Eq. \eqref{Ham3level_F-STIRAP}  was  diagonalized using the $\mathbf{T}(t)$ matrix in Eq. \eqref{Tmatrix} with the new Stokes field $\Tilde{\Omega}_{s_0}(t) = \Omega_{s1_0}(t)+\Omega_{s2_0}(t)e^{-i\phi}$  after imposing the condition that $\delta(t) =0$.  In Fig. \ref{C-F-STIRAP_dressed}, the pump and the Stokes Rabi frequency, the dressed state energies, the populations and coherence dynamics are plotted as a function of time for $\delta < 0$ and $\delta >0$. 
The population dynamics shows that it is possible to create the maximum coherence in the absence of the two-photon resonance by carefully choosing the chirp rates and chirping delay, satisfying the condition $\delta(t) =0$. The system is always aligned with the dark state $\ket{\lambda_0(t)}$, which is again given by the Eq. \ref{dark_state}, with the modified mixing angle $\theta(t)=\Omega_{p_0}(t)/|\tilde{\Omega}_{s_0}(t)|$. Owing to the modified Stokes field, as $t\rightarrow \infty$, $\theta(t)=\tan^{-1}[1] =\pi/4$, and the dark state now evolves from state $\ket{\tilde{1}}$ to $1/\sqrt{2}(\ket{\tilde{1}}-\ket{\tilde{3}})$, which is a maximally coherent superposition between the two states. The 'tilde' will again be dropped for convenience for the remainder of this section.

\begin{figure}
\includegraphics[scale=0.55]{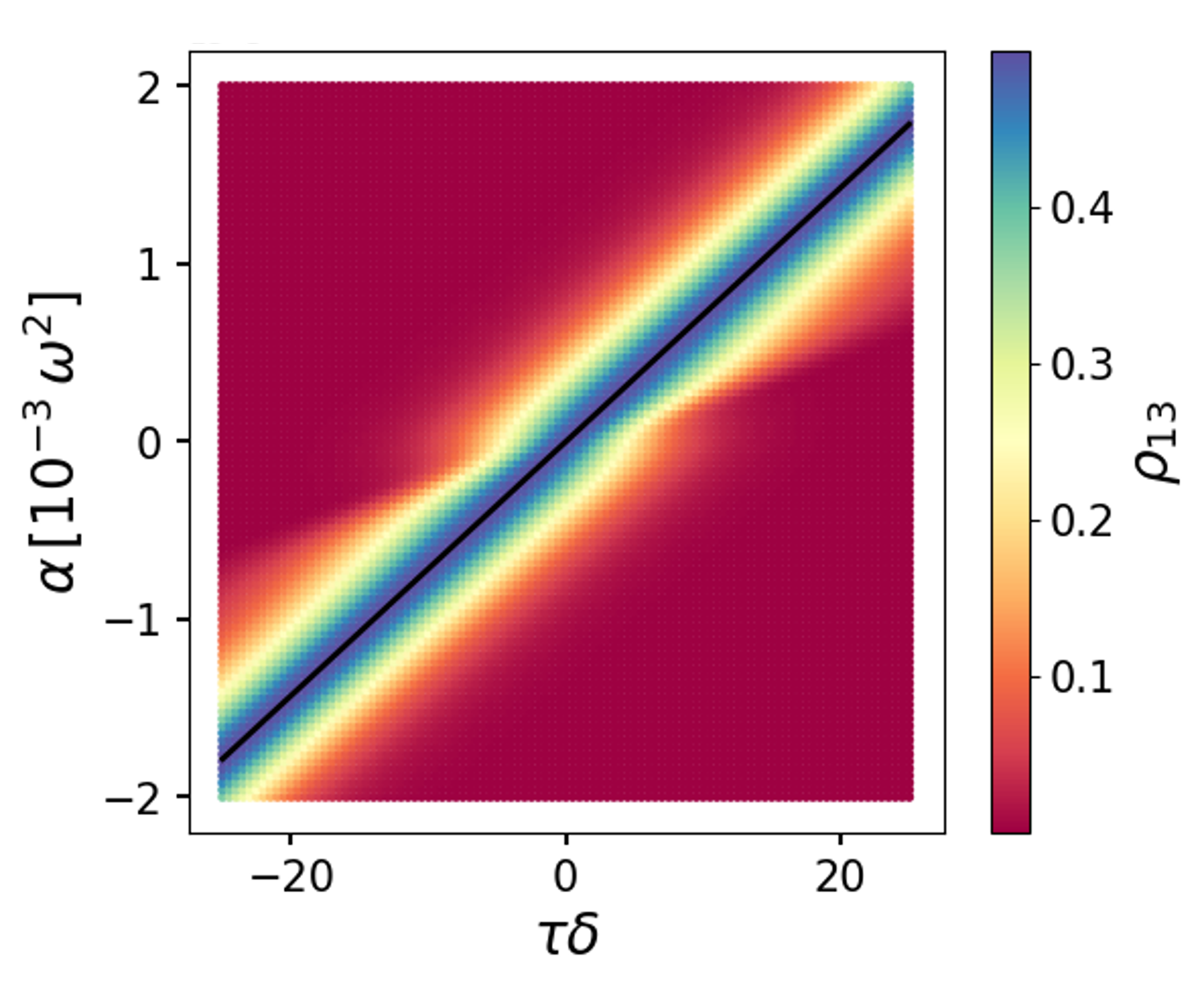}
\caption{Coherence in C-F-STIRAP as a function of two-photon detuning and chirp rate. The adiabatic regime and maximum coherence is achieved in the vicinity of the dark line satisfying $\alpha=\delta/2t_p$, corresponding to $\delta(t)=0$.}\label{coherence_delta_vs_chirp_3-level}
\end{figure}

The contour-plot of coherence between the initial and the final states, $\rho_{13}$, as a function of the two-photon detuning $\delta$ and the chirp rate $\alpha$ is shown in Fig.\ref{coherence_delta_vs_chirp_3-level}. The maximum coherence window, shown in blue color, is achieved  and remains relatively constant for the chirp values $\alpha=\delta / 2t_p$ satisfying the condition $\delta(t) =0$. This is in a stark contrast to the Fig. \ref{detuning_vs_coherence} in which state coherence decreased with the increase of the two-photon detuning. Thus, the delayed chirp in the C-F-STIRAP configuration overcomes the problem of maximizing state coherence adiabatically in the presence of the two-photon detuning.

\subsection{C-F-STIRAP in a  four-level $\lambda$ system with two energetically close final states}
\begin{figure}
\includegraphics[scale=0.6]{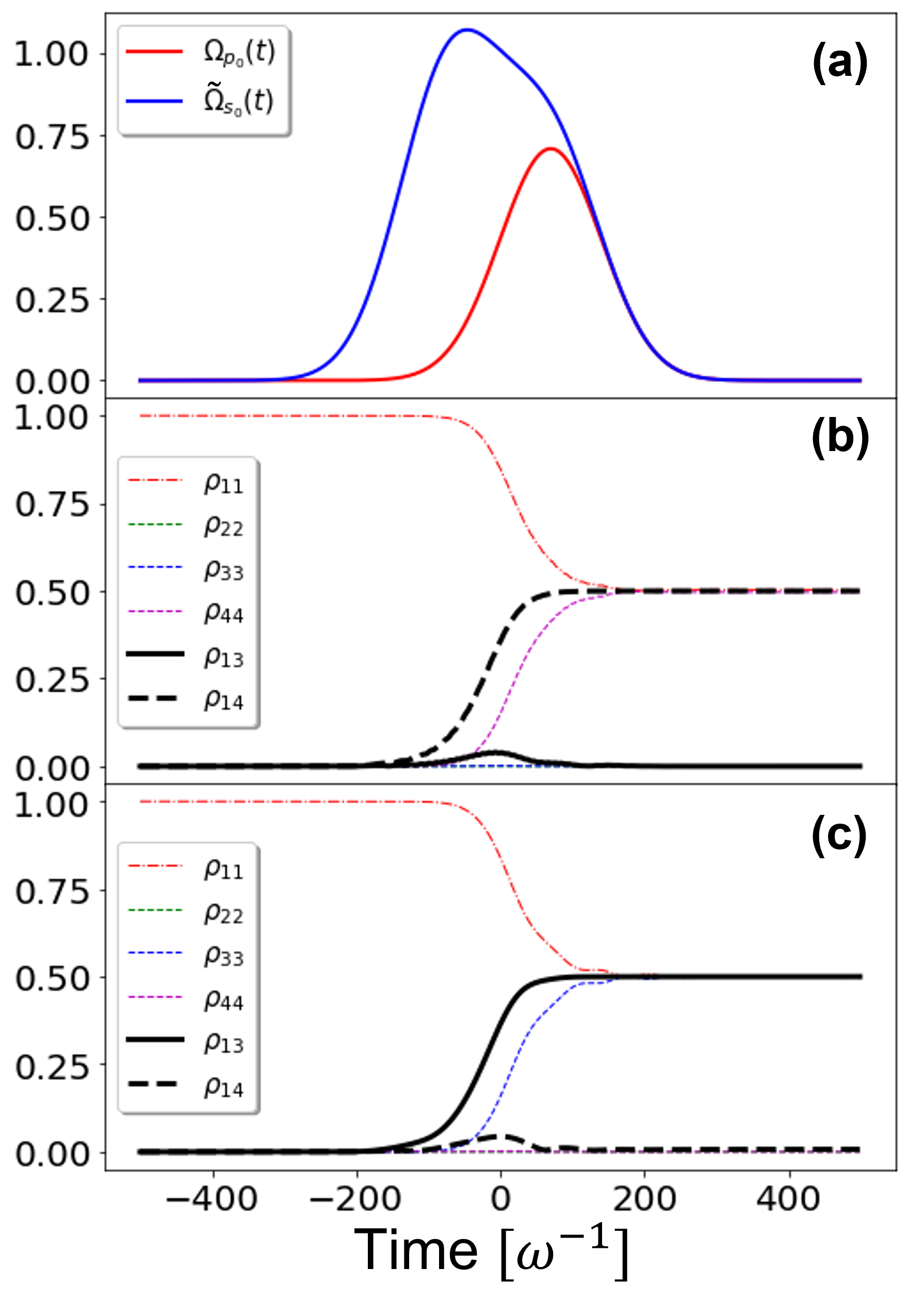}
\caption{The evolution of the Rabi frequencies (a), the population and coherence (b-c) in the case of chirped fractional STIRAP (C-F-STIRAP) for $\delta' = 0.14$. When the chirping delays are chosen to be $t_{d1}=0$ and $t_{d2}=-2t_p$, the coherence $\rho_{14}$ is maximized without populating state  $\ket{3}$, shown in (b). In contrast, when $t_{d1}=2t_p$ and $t_{d2}=0$ the states $\ket{1}$ and $\ket{3}$ receive the equal population and coherence $\rho_{13}$ is maximized without populating state $\ket{4}$, as shown in (c).}\label{4-level-F-STIRAP_coherence}
\end{figure}

In section IIC, it was shown that, using C-STIRAP scheme, the population can be driven fully to a desired level in a nearly degenerate four-level $\lambda$ system, Fig. \ref{4-level}. Motivated by this result we show that, using C-F-STIRAP, it is possible to create a system  with equal populations distributed between the initial and one of the desired final states in the four-level system.  In this case, the chirping of one of the Stokes pulses needs to be delayed for the selective final state excitation. The respective Hamiltonian in the field interaction frame for the four-level system is written by extending the three-level Hamiltonian in Eq. \eqref{Ham3level_F-STIRAP} to a four-level system as follows:
\begin{widetext}
\begin{eqnarray}
\mathbf{H}(t) &=& \frac{\hbar} {2}
\left( \begin{array}{cccc} 0       & \Omega_{p_0} (t)    &   0	&   0 \\
\Omega_{p0} (t) &  2[\Delta-\alpha(t-t_p)] &  \Omega_{s10}(t)+\Omega_{s20}(t)e^{i\eta(t)}   &   \Omega_{s10}(t)+\Omega_{s20}(t)e^{i\eta(t)} \\
0  &  \Omega_{s10}(t)+\Omega_{s20}(t)e^{-i\eta(t)} &  2(\delta'(t) + \delta') &   0\\
0  &  \Omega_{s10}(t)+\Omega_{s20}(t)e^{-i\eta(t)} & 0  &   2\delta'(t)
\end{array} \right), \nonumber \\
\, \nonumber \\
\label{Ham4level_C-F-STIRAP-1}
\end{eqnarray}
\end{widetext}
where,
\begin{gather}
\scalebox{0.9}{$\begin{aligned}
\delta'(t) &= -\delta'+\beta_1(t+t_p-t_{d1})-\alpha(t-t_p),\\ \, 
\eta(t)&= 2\omega_st_p+\frac{\beta_1}{2}(t+t_p-t_{d1})^2-\frac{\beta_2}{2}(t-t_p-t_{d2})^2.
\end{aligned}$}
\end{gather}
If the chirp rates are chosen to be $\alpha =\beta_1 =\beta_2 = \delta'/2t_p$, the fourth diagonal term becomes zero and the phase $\eta(t)$ becomes a constant, $\eta(t)=2\omega_st_p=\phi$, for the choice of chirping delays to be $t_{d1}=0$ and $t_{d2}=-2t_p$. 
This results in the creation of the maximum coherence between states $\ket{1}$ and $\ket{4}$, without populating any other states. The evolution of populations and coherence in this case is shown in Fig. \ref{4-level-F-STIRAP_coherence}(b). In a contrary scenario, a maximally coherent superposition is obtained between the initial and final, resonant state $\ket{3}$ when the chirp rates and chirping delays are chosen to be $\alpha =\beta_1 =\beta_2 = \delta'/2t_p$ and $t_{d1}=2t_p$ and $t_{d2}=0$. In this case, the third diagonal term becomes zero and the phase is again reduced to the same constant, $\eta(t)=2\omega_st_p=\phi$. The evolution of the state populations and coherence in this case are  shown in Fig. \ref{4-level-F-STIRAP_coherence}(c).

\begin{figure*}
\includegraphics[scale=0.56]{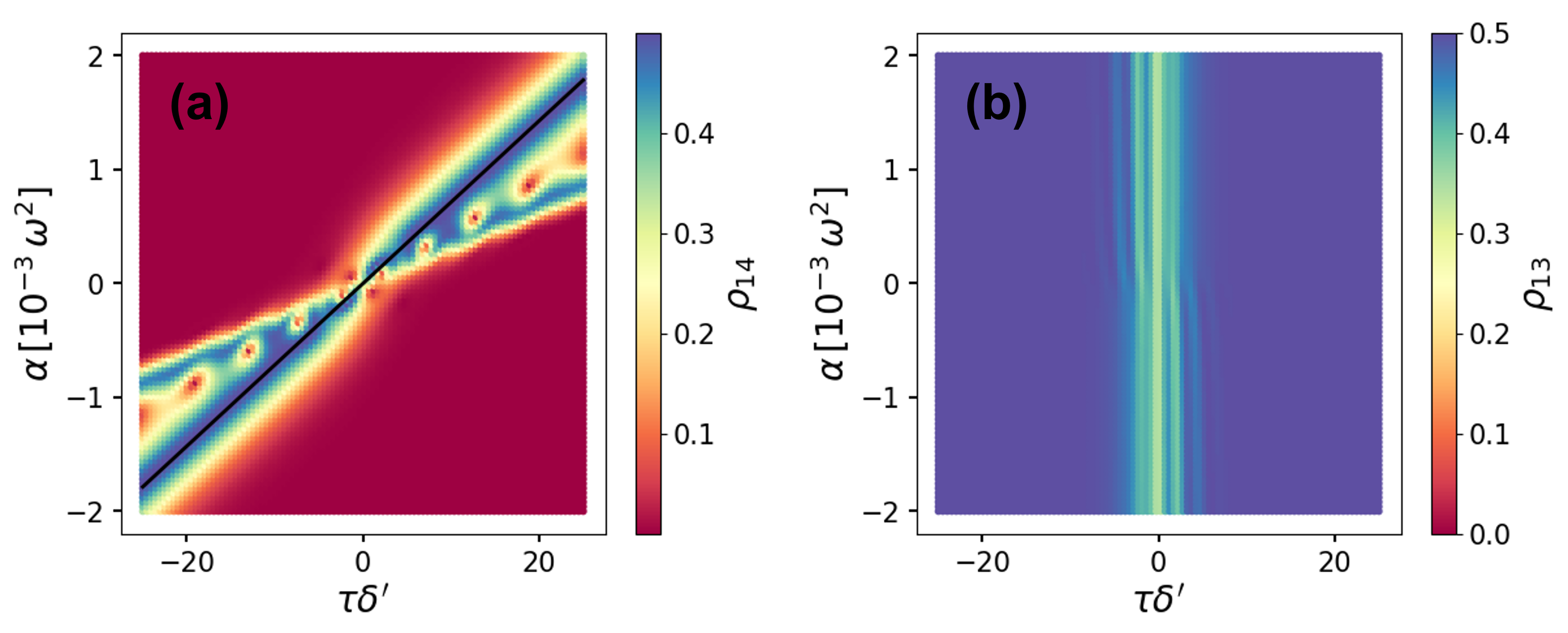}    
\caption{Coherence $\rho_{14}$ as a function of two-photon detuning and chirp rate $\alpha$ for a choice of chirping delays $t_{d1}=0$ and $t_{d2}=-2t_p$, (a), and $\rho_{13}$ for $t_{d1}=2t_p$ and $t_{d2}=0$, (b).  In (a), a maximum coherence between the initial and detuned state $\ket{4}$ is reached for chirp rates satisfying $\alpha =\delta'/2t_p$ represented by the diagonal dark line. As demonstrated in (b), there is no constraint condition on the chirp rate to create a maximally coherent superposition between the initial and resonant state $\ket{3}$ as long as the delays satisfy $t_{d1}=t_p$ and $t_{d2}=0$.}\label{4-level_coherence_contour}
\end{figure*}

\newpage
\subsection{Dressed state analysis of C-F-STIRAP in the four-level $\lambda$ system}

The total Stokes Rabi frequency is comprised of the effective Stokes Rabi frequency $\Omega_{s0}(t)$  and the effective Stokes phase $\theta_S(t)$, which are defined as follows:

\begin{widetext}

\begin{equation}
\begin{split}
\Omega_{s0}(t)&=\Omega_{s1_0}e^{-(t+t_p-t_{d1})^2/(2\tau^2)}\sqrt{1+\left(\dfrac{\Omega_{s2_0}}{\Omega_{s1_0}}e^{(2t_pt)/\tau^2}\right)^2+2\dfrac{\Omega_{s2_0}}{\Omega_{s1_0}}e^{(2t_pt)/\tau^2}\cos\left(\tilde{\beta}_1(t)-\tilde{\beta}_2(t)\right)}
\end{split},
\end{equation}
\end{widetext}

\begin{gather}
\scalebox{0.9}{$\begin{split}
&\tan(\theta_{s}(t))=\dfrac{\sin\left(\int\tilde{\beta}_1(t)\right)+\dfrac{\Omega_{s2_0}}{\Omega_{s1_0}}e^{(2t_pt)/\tau^2}\sin\left(\int\tilde{\beta}_2(t)\right)}{\cos\left(\int\tilde{\beta}_1(t)\right)+\dfrac{\Omega_{s2_0}}{\Omega_{s1_0}}e^{(2t_pt)/\tau^2}\cos\left(\int\tilde{\beta}_2(t)\right)}.
\end{split}$}
\end{gather}

The effective Stokes pulses chirped frequency, $g(t)$, is the derivative of the phase $\theta_S(t)$ and reads:
\begin{gather}
\scalebox{0.85}{$\begin{split}
g(t)&=\dfrac{d}{dt}\theta_s(t)=\dfrac{\Omega_{s1_0}(t)^2\tilde{\beta}_1(t)+\Omega_{s2_0}(t)^2\tilde{\beta}_2(t)}{\Omega_{s,0}(t)^2}\\
&+\Omega_{s1_0}(t)\Omega_{s2_0}(t)\dfrac{(\tilde{\beta}_1(t)+\tilde{\beta}_2(t))\cos\left(\int(\tilde{\beta}_1(t)-\tilde{\beta}_2(t))\right)}{\Omega_{s,0}(t)^2}\\
&-\Omega_{s1_0}(t)\Omega_{s2_0}(t)\dfrac{2t_p}{\tau^2}\dfrac{\sin\left(\int(\tilde{\beta}_1(t)-\tilde{\beta}_2(t))\right)}{\Omega_{s,0}(t)^2},
\end{split}$}
\end{gather}
where $\tilde{\beta}_1(t)=\beta_{1}(t+t_p-t_{d1})$ and $\tilde{\beta}_{2}(t)=\beta_{2}(t-t_p-t_{d2})$.

The transformation used to get to the field interaction representation is: 
\begin{equation}\label{51_trans_chirped}
\begin{aligned}
a_1(t) &= \tilde{a_1}(t)\\
a_2(t) &= \tilde{a_2}(t) e^{-i\left(\omega_p(t-t_p)+\tfrac{\alpha}{2}(t-t_p)^2\right)} \\
a_3(t) &= \tilde{a_3}(t) e^{i\left(\theta_s(t)-\omega_p(t-t_p)-\frac{\alpha}{2}(t-t_p)^2\right)}\\
a_4(t) &= \tilde{a_4}(t) e^{i\left(\theta_s(t)-\omega_p(t-t_p)-\frac{\alpha}{2}(t-t_p)^2\right)}.
\end{aligned}
\end{equation}

It leads to the field interaction Hamiltonian 

\begin{eqnarray}
\mathbf{H}(t) =
\frac{\hbar}{2} \left( \begin{array}{cccc}
0   & \Omega_{p0} (t) &   0 & 0\\
\Omega_{p0} (t) &  2\Delta(t) &  \Omega_{s0}(t) &  \Omega_{s0}(t)\\
0  &  \Omega_{s0}(t) &  2\delta'(t)+2\delta'& 0\\
0  &  \Omega_{s0}(t) & 0 & 2\delta'(t) \\
\end{array} \right), 
\label{Ham4level_F-STIRAP}
\end{eqnarray}
where $\Delta(t)=\Delta-\alpha(t-t_p)$, $\delta'(t)=  -\delta'(t)+g(t)-\alpha(t-t_p)$.


In F-STIRAP, the choice of the fields \eqref{F-STIRAP_envelopes}, gives us a different structure in the dressed state picture as compared to STIRAP. 
In the three-level system, the dressed states are non-degenerate at $t \rightarrow - \infty$ but the states $\ket{1}$, $\ket{3}$ are degenerate at $ t \rightarrow \infty$. 

\begin{figure*}
\centering
\includegraphics[scale=0.8]{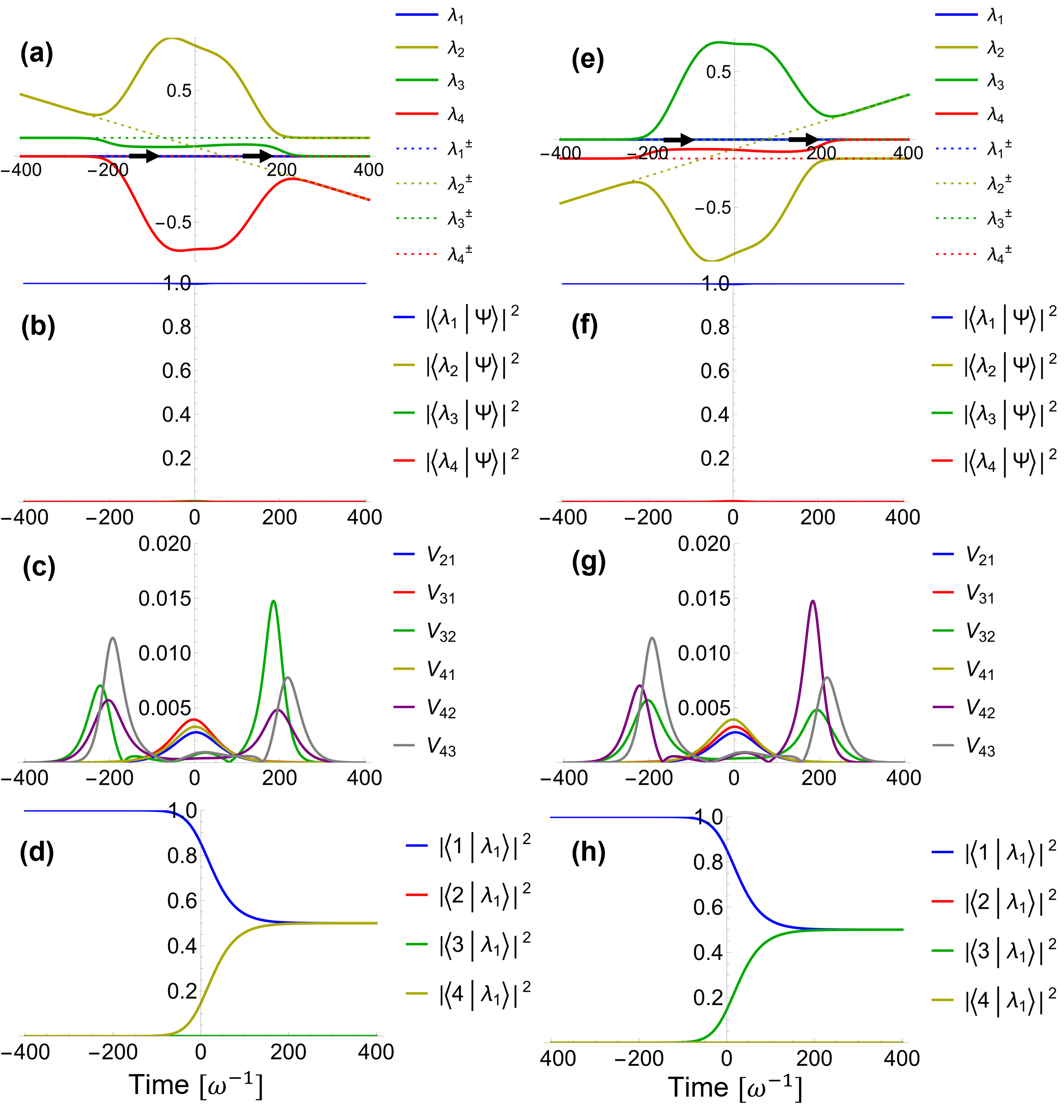}
\caption{Dressed state analysis 
of C-F-STIRAP four-level system. 
Left figures (a), (b), (c) and (d), correspond to the Fig. \ref{4-level-F-STIRAP_coherence}(b) when $\rho_{14}$ is maximized,  and right figures (e), (f), (g) and (h) correspond to the Fig. \ref{4-level-F-STIRAP_coherence}(c) when $\rho_{13}$ is maximized. 
The system is always aligned with the dressed state $\ket{\lambda_1}$ in both (a) and (e), as indicated by the arrows. Dressed state decomposition of wavefunction in (b) and (f)  confirms adiabatic evolution, with trivial non-adiabatic coupling rates, (c) and (g). 
The dressed state $\ket{\lambda_1}$ 
smoothly evolves to a maximum coherence state involving $\ket{1}$ and $\ket{4}$ in (d), and $\ket{1}$ and $\ket{3}$ in (h).
The parameters used are $\Delta=0$, $t_s=-70[\omega^{-1}]$, $t_p=70[\omega^{-1}]$, $ \tau_{p,s}=100[\omega^{-1}]$,   $\Omega_{{p_0,s_0}}=1.0[\omega]$, $\alpha = \pm1\times10^{-3} [\omega^2]$. 
}\label{4-leve_C-F-STIRAP_dressed_states}
\end{figure*}

The Schrodinger equation in the dressed state basis is  $\dot{\Psi}(t)=-i(E(t)+F(t))\Psi(t)$, where $E_{ij}(t)=(\lambda_i(t)+\braket{v_i(t)}{\dot{v}_i(t)})\delta_{ij}$ represents the diagonal adiabatic matrix, and $F_{ij}(t)=\braket{v_i(t)}{\dot{v}_j(t)}(1-\delta_{ij})$ represents the non-adiabatic coupling matrix. Non-adiabatic transitions need to be removed to achieve adiabatic passage, and this requires choosing parameters that prevent any avoided crossings and transitions between dressed states.

The time of avoided crossings between two dressed states can be found when the determinant of $H(t)$ and its derivative are both zero. There is also the possibility of a crossing with three or more dressed states. The probability of transitioning to the higher energy state $\ket{\lambda_{j}(t)}$, assuming all population is initially in state $\ket{\lambda_{i}(t)}$, during an avoided crossing at time $t$ where $\lambda_i(t)=\lambda_j(t)$, is given by the Landau-Zener formula \cite{DanileikoLZ3lvl}.

\begin{equation}
P_{LZ}(t)=\exp\left(-\dfrac{4\pi^2}{\hbar}\dfrac{\left(\braket{\lambda_{j}(t)}{\dfrac{d}{dt}\lambda_{i}(t)}\right)^2}{\dfrac{d}{dt}\left|\lambda_{i}(t)-\lambda_{j}(t)\right|}\right)\,.
\end{equation}
We note that while there can be many avoided crossings when the Rabi frequencies are small, these crossings trivially affect the bare state populations. The above discussions on the dressed states analysis are generic and are not specific to the case of C-F-STIRAP.

The energy gap between the two dressed states at $t \rightarrow \infty$, assuming we have no other non-adiabatic transitions, will determine the final state. The previous section gives us a scheme for which we can achieve adiabatic evolution to a maximum coherence state composed of equal population in the ground state and one of the final states, $\ket{3}$ or  $\ket{4}$. The selection of the latter state depends on  the chirp rate and frequency offsets, created by the introduced delays $t_{d1}$, $t_{d2}$ in the chirp functions. For adiabatic evolution, the energy spectrum of the Hamiltonian that satisfies this condition must be the one where the ground and the selected state coincide at the same energy at $t \rightarrow \infty$, and the non-selected state must diverge from the two previous states. This condition is required for the system to remain in a single dressed state.
For the choice of delays $t_{d1}$ and $t_{d2}$ used in Fig. \ref{4-level-F-STIRAP_coherence}, Hamiltonians \eqref{Ham4level_F-STIRAP} and \eqref{Ham4level_C-F-STIRAP-1} are the same, and the two field interaction pictures are equivalent. 

The analysis of the evolution of the dressed state  energies, shown in Fig. \ref{4-leve_C-F-STIRAP_dressed_states}, confirms that the creation of the maximum coherence in the four-level system via selective excitation in Fig. \ref{4-level-F-STIRAP_coherence} is perfectly adiabatic. The figures on the left, (a), (b), (c) and (d) correspond to the Fig. \ref{4-level-F-STIRAP_coherence}(b), where the delays are chosen to be $t_{d1}=0$ and $t_{d2}=-2t_p$, and the figures on the right correspond to the Fig. \ref{4-level-F-STIRAP_coherence}(c), where $t_{d1}=2t_p$ and $t_{d2}=0$. The system remains in the dressed (dark) state throughout the process, which smoothly evolves maximum superposition between $\ket{1}$ and $\ket{4}$ in (d) and between $\ket{1}$ and $\ket{3}$ in (h). The behavior of dressed states in (f), (g) and (h) are the same as (b), (c) and (d) respectively, except $\ket{\lambda_3}$ takes up the role of $\ket{\lambda_4}$ and vice versa. The rates of non-adiabatic couplings in (b) and (f) also confirm that the evolution is adiabatic as all of them have magnitudes much less than the difference between dressed energies.

\section{summary}

In this paper, we presented a novel scheme that selectively creates maximum coherence in a four-level system via chirped fractional stimulated Raman adiabatic passage (C-F-STIRAP). First, 
by analyzing the dressed state dynamics, we demonstrated that it is necessary to chirp the pulses in STIRAP in order to achieve adiabaticity in the absence of two-photon resonance. To eliminate the non-adiabatic contribution, both pulses must be chirped at the same rate and the value of two-photon detuning, $\delta$, must match the product of chirp rate and the time delay between the pulses, $\alpha(t_p-t_s)$. We then considered a four-level system with two nearly degenerate terminal levels and showed that the population can be driven exclusively to one of the terminal levels by the appropriate choice of the pulse chirping. For negative two-photon detuning, the detuned final state is populated if the chirp rate is positive and the resonant state is populated if it is negative. The constraint conditions on the chirp rate in both cases were discussed. The analysis of the evolution of dressed states revealed that the population transfer to the detuned state is adiabatic while the population transfer to the resonant state is non-adiabatic. Further, we showed that the population can be adiabatically driven to the resonant state by introducing a delay in the chirping of the Stokes pulse.

After the discussion of selective population transfer using C-STIRAP, we presented the theory of F-STIRAP and demonstrated that a maximally coherent superposition can be created in the three-level system in the presence of two-photon resonance. This was done by modifying the Stokes field as a superposition of two Stokes pulses thereby making both pump and Stokes fields vanish simultaneously. Later, we present the technique of C-F-STIRAP and showed that, by chirping all the pulses equally and introducing a chirping delay in the second Stokes pulse, the adiabatic creation of maximally coherent superposition is possible even in the absence of two-photon resonance. We then applied C-F-STIRAP technique to the previously considered four-level system and demonstrated that a maximal coherence between the initial and a pre-determined final state is possible by manipulating the chirping delays of Stokes pulses. The analysis of the evolution of dressed states confirms that the selective excitation in the four-level system is perfectly adiabatic owing to the choice of chirp rates and chirping delays.

Maximizing the coherence is crucial to optimizing the output signal in imaging and sensing techniques based on coherent Raman spectroscopy. Owing to the adiabaticity, robustness and higher spectral resolution, the method presented here may find various applications including in imaging and sensing methods. As a practical limitation, there is an upper limit on the value of two-photon detuning that can be compensated by the choice of chirp rate in both three-level and four-level systems. This is because of the limits on the values of possible temporal chirp rates for a given pulse duration and the requirement to have a significant overlap between the pulses for adiabatic passage. For a given pulse duration $\tau$, the time difference between the pump and Stokes pulses is taken as $t_p-t_s=1.4\tau$ in our calculations. For $\tau=100 [\omega^{-1}]$, the possible values of chirp rates lie between $\alpha \approx \pm 1\times 10^{-5} [\omega^2]$, implying that the two-photon detuning should be $|\delta|\lesssim |\alpha|(t_p-t_s)=0.007[\omega]$. In the three-level system, the detuning should not exceed this value to satisfy the adiabaticity condition and in four-level system, the separation of final levels should not exceed this value for selective excitation.



\section*{Acknowledgment}
S.M., J.Ch., and A.R. acknowledge support from the Office of Naval Research under awards
N00014-20-1-2086 and N00014-22-1-2374. S.M. acknowledges the Helmholtz Institute Mainz
Visitor Program and the Alexander von Humboldt Foundation.


{}
\end{document}